\documentclass[submission,copyright,creativecommons]{eptcs}
\providecommand{\event}{QAPL 2017} 
\usepackage{underscore}           

\usepackage{amsmath}
\usepackage{amsthm, amssymb, amsfonts}
\newtheorem{definition}{Definition}

\newtheorem{theorem}{Theorem}

\usepackage{graphicx}

\usepackage{xfrac}

\usepackage{enumitem}

\usepackage{subcaption}

\usepackage{scalerel}

\usepackage{enumitem}

\usepackage{relsize}
\usepackage[nounderscore]{syntax}
\usepackage{url}

\usepackage{array}
\usepackage{ifthen}
\usepackage{tikz}
\usetikzlibrary{fit,arrows,calc,decorations.markings,shapes.multipart}
\usepackage{listings}
\usepackage{wrapfig}
\usepackage{colortbl}

\newcolumntype{C}[1]{>{\centering\let\newline\\\arraybackslash\hspace{0pt}}m{#1}}

\tikzstyle{block} = [rectangle, draw, minimum width=0.5cm, minimum height=0.5cm]

\pgfdeclarelayer{background}
\pgfsetlayers{background,main}

\newcommand{\seml}[0]{\mathbb{[}\hspace{-1.5pt}\mathbb{[}}
\newcommand{\semr}[0]{\mathbb{]}\hspace{-1.5pt}\mathbb{]}}

\DeclareMathOperator*{\fuzzywedge}{\mathbf{\tilde{\wedge}}}
\DeclareMathOperator*{\fuzzyvee}{\mathbf{\tilde{\vee}}}
\DeclareMathOperator*{\fuzzyneg}{\mathbf{\tilde{\neg}}}

\DeclareMathSymbol{\mlq}{\mathord}{operators}{``}
\DeclareMathSymbol{\mrq}{\mathord}{operators}{`'}

\usepackage[ruled]{algorithm2e}

\title{Bridging Static and Dynamic Program Analysis \\
using Fuzzy Logic}
\author{Jacob Lidman \& Josef Svenningsson
\institute{Chalmers University of Technology}\email{\{lidman, josefs\}@chalmers.se}}

\def\titlerunning{Bridging Static and Dynamic Program Analysis using Fuzzy Logic}
\def\authorrunning{J. Lidman \& J. Svenningsson}
\begin{document}
\maketitle

\begin{abstract}
Static program analysis is used to summarize properties over all dynamic executions. In a unifying approach based on 3-valued logic properties are either assigned a definite value or unknown. But in summarizing a set of executions, a property is more accurately represented as being biased towards true, or towards false. Compilers use program analysis to determine benefit of an optimization. Since benefit (e.g., performance) is justified based on the common case understanding bias is essential in guiding the compiler. Furthermore, successful optimization also relies on understanding the quality of the information, i.e. the plausibility of the bias. If the quality of the static information is too low to form a decision we would like a mechanism that improves dynamically. 

We consider the problem of building such a reasoning framework and present the fuzzy data-flow analysis. Our approach generalize previous work that use 3-valued logic. We derive fuzzy extensions of data-flow analyses used by the lazy code motion optimization and unveil opportunities previous work would not detect due to limited expressiveness. Furthermore we show how the results of our analysis can be used in an adaptive classifier that improve as the application executes.
\end{abstract}

\section{Introduction}
How can one reconcile static and dynamic program analysis? These two
forms of analysis complement each other: static analysis summarizes
all possible runs of a program and thus provide soundness guarantees,
while dynamic analysis provides information about the particular runs
of a program which actually happen in practice and can therefore
provide more relevant information. Being able to combine these two
paradigms has applications on many forms on analyses, such as alias
analysis \cite{Mock01, Ribeiro07} and dependence
analysis \cite{Petersen96}.

Compilers use program analysis frameworks to prove \textit{legality} as well as determining \textit{benefit} of transformations. Specifications for legality are composed of \textit{safety} and \textit{liveness} assertions (i.e. universal and existentially quantified properties), while specifications for benefit use assertions that hold in the \textit{common case}. This reason for adopting the common case is that few transformations improve performance in general (i.e., for every input, environment). Similarly most transformations could potentially improve performance in a least one case. As such, compiler optimizations are instead motivated based on (an approximation of) the majority case, i.e. the (weighted) mean. While determining legality has improved due to advances in the verification community the progress in establishing benefit has been slow. 

In this paper we introduce fuzzy data-flow analysis, a framework for
static program analysis based on fuzzy logic. The salient feature of our
framework is that it can naturally incorporate dynamic information
while still being a static analysis. This ability comes thanks to a
shift from ``crisp'' sets where membership is binary, as employed in
conventional static analysis, to fuzzy sets where membership is
gradual.

\newpage
\noindent
We make the following contributions:
\begin{itemize}[noitemsep,topsep=0pt,parsep=0pt,partopsep=0pt]
 \item Section \ref{sec:fdfa} introduces our main contribution, the fuzzy data-flow framework.
 \item Section \ref{sec:lcm} demonstrates the benefit of our framework by presenting a generalization of a well-known code motion algorithm and we show how this generalization provides new opportunities for optimizations previous approaches would not discover.
 \item Section \ref{sec:lcm} shows how fuzzy logic can benefit program analysis by (1) using second-order fuzzy sets to separate uncertainty in data-flow and control-flow and hence improve an inter-procedural analysis and (2) using fuzzy regulators to refine the results of our static analysis, hence improving the precision dynamically.
\end{itemize}

\section{Preliminaries}
\label{sec:prelim}
We introduce and define fuzzy sets and the operators that form fuzzy logic. These concepts will be used in Section \ref{sec:fdfa} to define the transfer functions of our data-flow analysis.

\subsection{Fuzzy set}
Elements of a crisp set\footnote{In the context of fuzzy logic, crisp or Boolean set refer to a classical set to avoid confusion with fuzzy sets.} are either members or non-members w.r.t to a universe of discourse. A fuzzy set (FS) instead allow partial membership denoted by a number from the unit interval $[0,1]$. The membership degree typically denotes vagueness. The process to convert crisp membership to fuzzy grades is called \textit{fuzzification} and the inverse is called \textit{defuzzification}.
Following Dubois et al.~\cite{dubois2000fundamentals,DuboisPrade80} let $S$ be a crisp set and $\mu: S \mapsto [0,1]$ a \textit{membership function} (MF) then $\left<S, \mu\right>$ is a fuzzy set. As a convention, if $S$ is understood from context we sometimes refer to $\mu$ as a fuzzy set. The membership function formalizes the fuzzification. 
%
%
%
%
Fuzzy sets are ordered point-wise, i.e. $(S,\mu_A) \leq (S,\mu_B) \Leftrightarrow \forall s \in S\colon \mu_A(s) \leq \mu_B(s)$.  

We can accommodate some notion about uncertainty of vagueness by considering a type-2 fuzzy set where the membership degree itself is a fuzzy set.
Type-2 FS (T2FS) membership functions are composed of a primary ($J_s$) and secondary ($\mu$) membership $\{\left<(s,u), \mu(s,u)\right> \mid s \in S, u \in J_s \subseteq [0,1]\}$. Here uncertainty is represented by the secondary membership that define the possibility of the primary membership. When for each $x$ and $u$, it holds $\mu(x,u) = 1$ the T2FS is called an \textit{interval} T2FS. Gehrke et al.~\cite{Gehrke96} showed that this can equivalently be described as an interval valued fuzzy sets (IVFS) where $\mu\colon S \rightarrow \{[l,u]\left| \bot \leq l \leq u \leq \top\right.\}$. IVFS are a special case of lattice valued fuzzy sets ($L$-fuzzy sets) where the membership domain forms a lattice over $[0,1]$. Defuzzification of T2FS often proceeds in two phases. The first phase applies \textit{type reduction} to transform the T2FS to a type-1 FS (T1FS). The second phase then applies a type-1 defuzzification. 

\subsection{Fuzzy logic}
Fuzzy logic defines many-valued formal systems to reason about truth in the presence of vagueness. Contrary to classical logic the law of excluded middle ($p \vee \lnot p = \top$) and the law of non-contradiction ($p \wedge \lnot p = \bot$) does not, in general, hold for these systems. Fuzzy logic uses T-, S- and C- norms to generalize the logical operators $\wedge$, $\vee$ and $\neg$. We compactly represent a fuzzy logic by $\left<\fuzzywedge, \fuzzyvee, \fuzzyneg\right>$\footnote{Although one would expect the definition of a fuzzy logic to include a ``fuzzy implication'' operator in this work we do not consider it.} which is sometimes called a \textit{De Morgan system} \cite{dubois2000fundamentals} because it satisfies a generalization of De Morgans laws: $\fuzzyneg(P \fuzzywedge Q) \Leftrightarrow \fuzzyneg P \fuzzyvee \fuzzyneg Q$ and $\fuzzyneg(P \fuzzyvee Q) \Leftrightarrow \fuzzyneg P \fuzzywedge \fuzzyneg Q$.

\begin{definition} Let $U$ be a binary function $[0,1]^2 \to [0,1]$ that is commutative, associative and increasing and has an identity element $e \in [0,1]$. If $e = 1$ then $U$ is a \textbf{Triangular norm (T-norm)} and if $e = 0$ then $U$ is a \textbf{Triangular conorm (S-norm)}\footnote{The general concept, allowing any $e \in [0,1]$, is called a \textit{uninorm} \cite{dubois2000fundamentals} and is either orlike (i.e., $U(0,1) = U(1,0) = 1$) or andlike (i.e., $U(0,1) = U(1,0) = 0$). Our work does not require the full generality.}.
\end{definition}
\begin{definition} A \textbf{C-norm} is a unary function $n\colon [0,1] \to [0,1]$ that is decreasing, involutory (i.e., $n(n(x)) = x$) with boundary conditions (i.e, $n(0) = 1, n(1) = 0$).
\end{definition}
\noindent
Standard examples of fuzzy logics are shown in Table \ref{tbl:fl} \cite{dubois2000fundamentals,DuboisPrade80}. Examples 1-3 are special cases (and limits) of the Frank family of fuzzy logics that are central to our work and formally defined in Definition \ref{defFrankFamily}.

\begin{table}
	\center
	\begin{tabular}{l|l|l|l|l}
		& \textbf{Fuzzy logic} & \textbf{T-norm} & \textbf{S-norm} & \textbf{C-norm} \\ \hline
 		\textbf{1} & \textit{Min-Max} & $\min(x,y)$ & $\max(x,y)$ & $1-x$ \\
 		\textbf{2} & \textit{Algebraic Sum-product} & $xy$ & $x+y-xy$ & $1-x$ \\ 
 		\textbf{3} & \textit{Lukasiewicz} & $max(x+y-1, 0)$ & $min(x+y, 1)$ & $1-x$ \\
 		\textbf{4} & \textit{Nilpotent} & $\begin{cases} min(x,y) & x+y > 1 \\ 0 & \text{otherwise}\end{cases}$ & $\begin{cases} max(x,y) & x+y < 1 \\ 1 & \text{otherwise}\end{cases}$ & $1-x$ \\ \hline	
	\end{tabular}
	\caption{Common instantiations of fuzzy logics}
	\label{tbl:fl}	
	\vspace*{-0.75cm}
\end{table}
\begin{definition}\label{defFrankFamily} Let $s \in [0,1] \cup \{\infty\}$ then the \textbf{Frank family} of T-norms is defined by: 

$T^s(x,y) = \begin{cases}
		\min(x,y) & s = 0 \\
		xy & s = 1 \\
		\max(x+y-1,0) & s = \infty \\
		log_s\left(1 + \frac{\left(s^x - 1\right)\left(s^y - 1\right)}{s-1}\right) &
			\text{otherwise}\end{cases}$
\end{definition}

\noindent
The set of intervals in $[0,1]$ forms a bounded partial order $\left<\mathbb{I}, \sqsubseteq, \top, \bot\right>$\footnote{This should not be confused with the partial order used in the interval abstraction.} where $[l_x, u_x] \leq [l_y, u_y] \Leftrightarrow \left(l_x \leq l_y\right) \\ \wedge \left(u_x \leq u_y\right), \top = [1, 1]$ and $\bot = [0,0]$. 
As per Gehrke et al.~\cite{Gehrke96} we can point-wise lift a T1FS fuzzy logic $\left<\fuzzywedge, \fuzzyvee, \fuzzyneg\right>$ to a IVFS fuzzy logic, i.e., $[l_x, u_x] \odot [l_y, u_y] = [l_x \odot l_y, u_x \odot u_y], \hat{\odot} \in \{\fuzzywedge, \fuzzyvee\}$ and $\fuzzyneg [l, u] = [\fuzzyneg u, \fuzzyneg l]$.

\section{Fuzzy data-flow analysis}
\label{sec:fdfa}
Static data-flow analyses deduce values of semantic properties that are satisfied the dynamics of the application. The dynamics is formalized as a system of monotone transfer functions and collector functions. Transfer functions describe how blocks alter the semantic properties. Collectors functions merge results from different, possibly mutual exclusive, branches of the application. The solution of the analysis is obtained through Kleene iteration and is a unique fixed-point of the system of equations. In a classical framework the domain of the values is binary, i.e. either true (1) or false (0). The interpretation of these values depends on the type of analysis. The value true means that the property can possibly hold in a \textit{may}-analysis (i.e., it is impossible that the value is always false) while it means that the property always holds in a \textit{must}-analysis. The value false could mean either the opposite of true or that the result is inconclusive.

Our fuzzy data-flow analysis instead computes the partial truth of the property, i.e. values are elements of $[0,1]$. A value closer to $0$ means that the property is biased towards false and vice versa. Furthermore the transfer functions are logical formulas from a Frank family fuzzy logic and the collector functions are weighted average functions where the constant $\alpha$ is determined prior to performing the analysis. In contrast to the classical framework Kleene iteration proceeds until the results differ by a constant $\epsilon$ which is the maximal error allowed by the solution. The error can be made arbitrarily small. 

\begin{figure}[!ht]
	\begin{minipage}{0.3\textwidth}
	  \scriptsize
	  \begin{tikzpicture}
		 \node[block] (f0) {$Out = 0.0$};
		 	\node[left of=f0,xshift=-0.25cm] (f0Name) {B0};
		 \node[block, below of=f0] (f1) {$Out = In$};
		 	\node[left of=f1,xshift=-0.15cm] (f1Name) {B1};		 
		 \node[block, below left of=f1, xshift=-0.5cm,yshift=-1.5cm] (f2) {$Out = 0.8 \wedge \left(\lnot In \vee \lnot 0.7\right)$};
		 	\node[below of=f2,yshift=0.4cm] (f1Name) {B2};
		 \node[block, below right of=f1, xshift=0.5cm, yshift=-0.5cm] (f3) {$Out = In$};
		 	\node[below of=f3,yshift=0.5cm,xshift=0.25cm] (f1Name) {B3};
		 \draw[->] (f0) node[left,xshift=0cm,yshift=-0.5cm,text=red]{$0.1$} -- (f1);
		 \path[draw,->] node[left,xshift=0.4cm,yshift=-2.25cm,text=red]{$1.0$} ($(f1.south)+(0.1,0)$) to [bend left] ($(f2.north)+(0.1,0)$);
		 \path[draw,->] node[right,xshift=-1.75cm,yshift=-2.25cm,text=red]{$0.9$} (f2.north) to[bend left] (f1.south);
		 \draw[->] (f1) node[right,xshift=0.5cm,yshift=-0.6cm,text=red]{$1.0$} -- (f3);
	  \end{tikzpicture}	
	\end{minipage}%
	\hspace*{0.01cm}\vrule\hspace*{0.01cm}
	\begin{minipage}{0.33\textwidth}
	  \scriptsize
	  $\begin{cases}
	  	Out(B0) = 0.0 \\
	  	Out(B1) = \textcolor{red}{0.1Out(B0) + 0.9Out(B2)} \\
	  	Out(B2) = \min\left(0.8, \max(1-\textcolor{red}{Out(B1)}, 0.3)\right) \\
	  	Out(B3) = \textcolor{red}{Out(B1)}
	  \end{cases}$
	\end{minipage}%
	\vrule\hspace*{0.01cm}
	\begin{minipage}{0.38\textwidth}
	  \includegraphics[width=\textwidth]{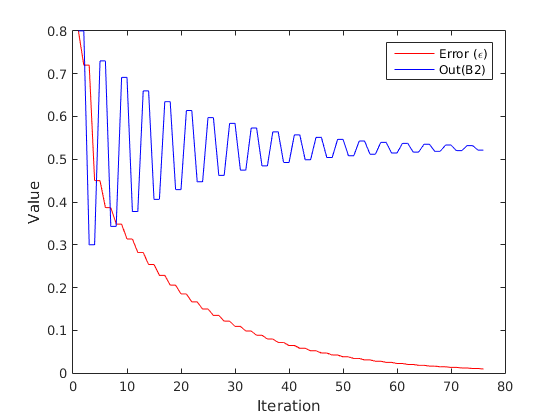} 
	\end{minipage}%
 \caption{Example flow-graph (left) and its corresponding equation system (middle) and the analysis result and error as a function of iteration (right)}
 \label{fig:Example}
\end{figure}
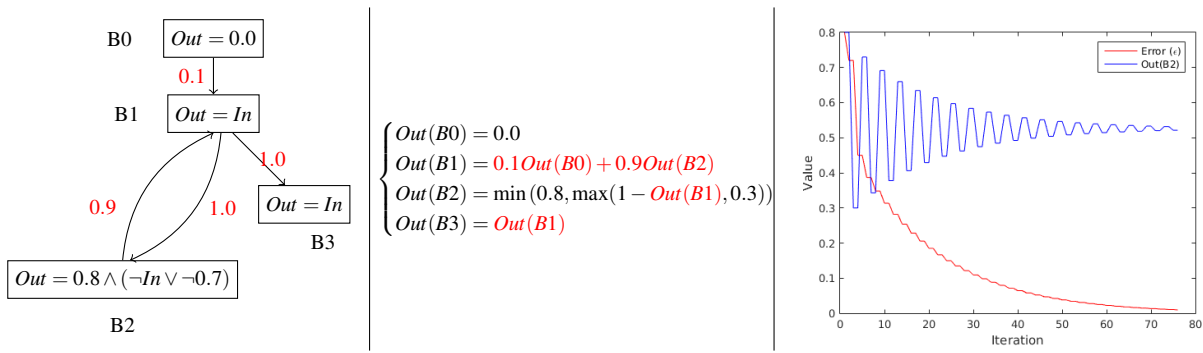

This section introduces the fuzzy data-flow framework and we prove termination using continuity properties and Banach's fixed-point theorem. Section \ref{sec:lcm} then presents an example analysis to demonstrate the benefit of the framework. The analysis is performed on a weighted flow-graph $G = \left<V, E, \alpha\right>$ where $V$ is a set of logical formulas (denoting the transfer function of each block), $E \subseteq V \times V$ is a set of edges (denoting control transfers) and $\alpha_e \in [0,1]$ denotes the normalized contribution for each edge $e$. As a running example we will use Figure \ref{fig:Example} (left) which shows a flow-graph with four nodes and their corresponding logical formula. The flow graph has four control edges denoting contributions between nodes. For instance, Block 1 (B1) receives 0.1 of its contribution from B0 and 0.9 from B2, i.e. $\alpha_{\left<B0,B1\right>} = 0.1$ and $\alpha_{\left<B2,B1\right>} = 0.9$.

\begin{definition}\label{defFDFA} Let $\mathcal{P}$ be a finite set of properties and $VS\colon \mathcal{P} \mapsto [0,1]$ a valuation for each property. We use $\seml \phi \semr \left(VS\right)$ to denote the interpretation of the fuzzy formula $\phi$ given a $VS$. Given a flow-graph $G = \left<V, E, \alpha\right>$ with a unique start node $v_\mathit{start}$ the map $GS\colon V \mapsto VS$ describes the value of each property at each node and a fuzzy data-flow analysis is a Kleene iteration of $F\colon GS \mapsto GS$:

$F(S) = \lambda v. \begin{cases}
			S(v_{start}) & v = v_\mathit{start} \\
			\sum\limits_{\left<w,v\right> \in E} \alpha_{\left<w,v\right>}\seml v \semr \left(S(w)\right) & \text{otherwise} 
			\end{cases}$
\end{definition}

\noindent
Figure \ref{fig:Example} (middle) shows the equation system, as implied by Definition \ref{defFDFA}, interpreted in a min-max fuzzy logic  for the example flow-graph. The red colored text corresponds to the collector function, i.e. the weighted average, and the normal text is the interpretation of the logical formula. In order to prove termination of a fuzzy analysis we need to introduce a continuity property. 
\begin{definition} \label{defLipFunc} A function $f\colon [0,1]^n \mapsto [0,1]$ is $K$-\textbf{Lipschitz continuous}\footnote{Our definition restricts the domain and metric of both metric spaces (i.e., for the domain and co-domain of $f$) compared to the more general, and common, definition of a Lipschitz continuous function.} iff $\forall x,h\colon |f(\vec{x}-\vec{h}) - f(\vec{x})|_1 \leq K|\vec{h}|_1$. Where $|\vec{x}|_1$ is $l_1$-norm (i.e., the absolute value) of $\vec{x}$\footnote{Other $l_p$-norms can be used but only if we restrict the logic to the min-max fuzzy logic \cite{Mesiarová07}.}. If $0 \leq K < 1$ then $f$ is called a \textbf{contraction mapping} and if $0 \leq K \leq 1$ then $f$ is called a \textbf{non-expansive mapping}.
\end{definition}
In a sequence of applications of a contraction mapping the difference between two consecutive applications will decrease and in the limit reach zero. By imposing a bounded error we guarantee that this sequence terminates in a bounded amount of time. The analysis error and result of B2 as a function of iteration for the example is shown in Figure \ref{fig:Example} (right). Note that the error (red line) is decreasing and the value of B2 (blue line) tends towards a final value.
We next proceed to prove that any fuzzy data-flow analysis iteratively computes more precise results and terminates in a bounded amount of time for a finite maximum error $\frac{1}{2^q}$ from some $q \in \mathbb{N}-\{0\}$. We let $[0,1]_q$ denote the maximal congruence set of elements from $[0,1]$ that are at least $\frac{1}{2^q}$ apart, i.e. $[0,1]_q = \{\frac{i}{2^q} \mid 0 \leq i \leq 2^q\}$. The set of intervals on $[0,1]$, i.e. $\mathbb{I}$ are defined analogously. For this we prove the \textit{non-expansive} property of fuzzy formulas.

\begin{theorem} \label{theoremLipFunc} Let $x,y,C,w_i \in [0,1]_q$, for some $i \in \mathbb{N}$, $f_i(\vec{x})\colon [0,1]_q^n \mapsto [0,1]_q$ be $1$-Lipschitz and\\ $g_i(\vec{x})\colon  [0,1]_q^n \mapsto [0,1]_q$ be $K_i$-Lipschitz.
 \begin{itemize}[noitemsep,topsep=0pt,parsep=0pt,partopsep=0pt]
  \item Functions $x+y$, $x-y$, $xy$, $\min(x,y)$ and $abs(x)$ are $1$-Lipschitz. Constants are $0$-Lipschitz.
  \item If $\sum_{i = 0}^{N-1} w_i = 1$ then $\sum_{i = 0}^{N-1} w_if_i(\vec{x})$ is $1$-Lipschitz.
  \item The composition $g_a \circ g_b$ is $K_aK_b$-Lipschitz.
 \end{itemize}
Finally,
 \begin{itemize}[noitemsep,topsep=0pt,parsep=0pt,partopsep=0pt]
   \item Formulas defined in a Frank family Fuzzy logic are $1$-Lipschitz.
   \item If $F\colon \mathbb{I}_q^n \to \mathbb{I}_q$ satisfies $\forall x \in \mathbb{I}_q^n\colon y \in x \Rightarrow f(y) \in F(x)$ then F is $1$-Lipschitz.
 \end{itemize}
\end{theorem}

\noindent
In summary, as per Theorem \ref{theoremLipFunc}:
\begin{itemize}[noitemsep,topsep=0pt,parsep=0pt,partopsep=0pt]
 \item Transfer functions in a Frank family fuzzy logic are non-expansive mappings.
 \item $S(v_\mathit{start})$ is constant and hence a contraction mapping.
 \item The composition of 1) Two non-expansive functions is a non-expansive function and 2) A non-expansive and a contraction function is a contraction function.
\end{itemize}
\noindent
As the analysis is performed on the unit interval which together with the $l_1$-norm forms a complete metric space we can guarantee termination by Banach's fixed-point theorem.

\begin{theorem}[Banach fixed-point theorem] Let $\left(X, d\right)$ be a complete metric space and $f\colon X \mapsto X$ a contraction. Then $f$ has a unique fixed-point $x^*$ in $X$.
\end{theorem}

\noindent
This concludes our development of fuzzy data-flow analysis.

\section{Lazy code motion}
\label{sec:lcm}
Improving performance often means removing redundant computations. Computations are said to be fully redundant, but not dead, if the operands at all points remain the same. For two such computations it is enough to keep one and store away the result for later. We can eliminate this redundancy using (global) common sub-expression elimination (GCSE). Furthermore a computation that does not change on some paths is said to be partially redundant. Loop invariant code motion (LICM) finds partially redundant computations inside loops and move these to the entry block of the loop. 
Lazy code motion is a compiler optimization that eliminate both fully and partially redundant computations, and hence subsumes both CSE and LICM. Knoop-R\"{u}thing-Steffen (KRS) algorithm \cite{Knoop1992, Drechsler1993} performs LCM in production compilers such as GCC when optimizing for speed. 
\newpage
\noindent
It consists of a series of data-flow analyses and can be summarized in these four steps:
\begin{enumerate}[noitemsep,topsep=0pt,parsep=0pt,partopsep=0pt]
 \item Solve a very busy expression\footnote{Knoop et al.~\cite{Knoop1992} refer to this as anticipatable expression data-flow problem.} and an available expression data-flow problem~\cite{Nielson1999}.
 \item Introduce a set that describes the earliest block where an expression must be evaluated.
 \item Determine the latest control flow edge where the expression must be computed.
 \item Introduce $Insert$ and $Delete$ sets which describe where expressions should be evaluated.
\end{enumerate}

\noindent
The target domain of the analysis is the set of static expressions in a program. Input to the analysis is three predicates determining properties about the expressions in different blocks:
\begin{itemize}[noitemsep,topsep=0pt,parsep=0pt,partopsep=0pt]
 \item An expression ``$e$'' is \textit{downward exposed} if it produces the same result if evaluated at the end of the block where it is defined. We use $DEE(b,e)$ to denote if ``$e$'' is downward exposed in block ``$b$''.
 \item An expression ``$e$'' is \textit{upward exposed} if it produces the same result if evaluated at the start of the block where it is defined. We use $UEE(b,e)$ to denote this.
 \item An expression ``$e$'' is \textit{killed} in block ``$b$'' if any variable appearing in ``$e$'' is updated in ``$b$''. We use $KILL(b,e)$ to denote this.
\end{itemize}

\begin{figure}
\scriptsize
	\begin{tikzpicture}[scale=0.87, every node/.style={scale=0.87}]
		
		\node (fDesc) {\normalsize \textbf{Knoop-Ruthing-Steffen LCM}}; 
		\node[block, below of=fDesc, yshift=-0.5cm] (f1) {\footnotesize
	     	\begin{tabular}{ll}		 
			 	$\begin{array}{l} \textbf{Available} \\ \textbf{expression} \end{array}$ & $\begin{cases} 
					AvIn(b) = \bigwedge_{b' \in Pred(b)} AvOut(b'), b \neq B0 \\
					AvOut(b) = DEE(b) \vee \left[AvIn(b) \wedge \lnot Kill(b)\right]
				\end{cases}$ \\
				\\			
				$\begin{array}{l} \textbf{Very busy} \\ \textbf{expression} \end{array}$ & $\begin{cases}
					AnIn(b) = \bigwedge_{b' \in Succ(b)} AnOut(b'), b \neq B5 \\
					AnOut(b) = UEE(b) \vee \left[AnIn(b) \wedge \lnot Kill(b)\right]
				\end{cases}$
			\end{tabular}};	
		 \node[left of=f1, xshift=-3.9cm] (f1Desc) {\normalsize (1)};
			
		 \node[block, below of=f1, yshift=-1.5cm] (f2) {\footnotesize
		 	$Earliest(i,j) = \begin{cases} AnIn(j) \wedge \lnot AvOut(i) \wedge \left[Kill(i) \vee AnOut(i)\right] & i \neq B0 \\ AnIn(j) \wedge \lnot AvOut(B0) & \text{otherwise}\end{cases}$};
	 	 \node[left of=f2, xshift=-4.25cm] (f2Desc) {\normalsize (2)};	
		 	
		 \node[block, below of=f2, yshift=-1cm] (f3) {\footnotesize
			 	$\begin{cases}
					LaterIn(j) = \bigwedge_{j' \in Pred(j)} LaterOut(:,j), j \neq B0 \\
					LaterOut(i,j) = Earliest(i,j) \vee \left[LaterIn(i) \wedge \lnot UEE(i)\right]
				\end{cases}$};
	 	 \node[left of=f3, xshift=-3.25cm] (f3Desc) {\normalsize (3)};				
				
		 \node[block, below of=f3, yshift=-1cm] (f4) {\footnotesize
		 	\begin{tabular}{l}
		 	 $Insert(i,j) = LaterOut(j) \wedge \lnot LaterIn(j)$ \\
		 	 $Delete(k) = UEE(k) \wedge \lnot LaterIn(k), k \neq B0$
		 	\end{tabular}};	
	 	 \node[left of=f4, xshift=-2.45cm] (f4Desc) {\normalsize (4)};

	    
	    	     	
	     \node[block, fill=gray!20, right of=fDesc, yshift=-1cm, xshift=7.5cm] (fInitClas) {
	     	\begin{tabular}{l|lllllll ll}
	     	 \textbf{Block} & \textbf{DEE} & \textbf{UEE} & \textbf{KILL} \\
	     	 	& 6543210 & 6543210 & 6543210\\ \hline
	     	 B0 & 0000000 & 0000000 & 1111111 \\
	     	 B1 & 0000001 & 0000001 & 0000000 \\
	     	 B2 & 0000010 & 0000010 & 0000000 \\
	     	 B3 & 0000000 & 1000000 & 1100010 \\
	     	 B4 & 0101000 & 0110100 & 1011111 \\
	     	 B5 & 0000000 & 0000000 & 0000000
	     	\end{tabular}};    	
	     	
	     \node[block, fill=gray!20, below of=fInitClas, xshift=0cm, yshift=-11.5cm] (fExitClas) {
	     	\begin{tabular}{ll|ll}
	     	 \textbf{Edge} & \textbf{Insert} & \textbf{Block} &\textbf{Delete} \\
	     	 				   & 6543210 & & 6543210 \\ \hline
	     	 B0$\rightarrow$B1 & 0000000 & \textbf{B0} & 0000000 \\
	     	 B1$\rightarrow$B5 & 0000000 & \textbf{B1} & 0000000 \\
	     	 B1$\rightarrow$B3 & 0000000 & \textbf{B2} & 0000000 \\
	     	 B2$\rightarrow$B3 & 0000000 & \textbf{B3} & 0000000 \\
	     	 B2$\rightarrow$B4 & 0000000 & \textbf{B4} & 0000000 \\
	     	 B3$\rightarrow$B4 & 0000000 & \textbf{B5} & 0000000 \\
	     	 B4$\rightarrow$B1 & 0000000 & & \\
	     	\end{tabular}};

	    
			    	
	     \node[block, fill=gray!80, below of=fInitClas, xshift=0.75cm, yshift=-5cm] (fInitFuzzy) {
	     	\begin{tabular}{l||l|l|l|l|l|l|l|}
	     	 \textbf{Block} & \multicolumn{7}{c}{$DEE$} \\
	     	 	& 6 & 5 & 4 & 3 & 2 & 1 & 0 \\ \hline
	     	 B0 & 0.0 & 0.0 & 0.0 & 0.0 & 0.0 & 0.0 & 0.0 \\
	     	 B1 & 0.0 & 0.0 & 0.0 & 0.0 & 0.0 & 0.0 & 1.0 \\
	     	 B2 & 0.0 & 0.0 & 0.0 & 0.0 & 0.0 & 1.0 & 0.0 \\
	     	 B3 & 0.0 & 0.0 & 0.0 & 0.0 & 0.0 & 0.0 & 0.0 \\
	     	 B4 & 0.0 & 1.0 & 0.0 & 1.0 & 0.0 & 0.0 & 0.0 \\
	     	 B5 & 0.0 & 0.0 & 0.0 & 0.0 & 0.0 & 0.0 & 0.0 \\
	     	 \\
			\textbf{Block} & \multicolumn{7}{c}{$UEE$} \\
				& 6 & 5 & 4 & 3 & 2 & 1 & 0 \\ \hline
	     	 B0 & 0.0 & 0.0 & 0.0 & 0.0 & 0.0 & 0.0 & 0.0 \\
	     	 B1 & 0.0 & 0.0 & 0.0 & 0.0 & 0.0 & 0.0 & 1.0 \\
	     	 B2 & 0.0 & 0.0 & 0.0 & 0.0 & 0.0 & 1.0 & 0.0 \\
	     	 B3 & 1.0 & 0.0 & 0.0 & 0.0 & 0.0 & 0.0 & 0.0 \\
	     	 B4 & 0.0 & 1.0 & 1.0 & 0.0 & 1.0 & 0.0 & 0.0 \\
	     	 B5 & 0.0 & 0.0 & 0.0 & 0.0 & 0.0 & 0.0 & 0.0 \\
	     	 \\
	     	 \textbf{Block} & \multicolumn{7}{c}{$KILL$} \\ 
	     	 	& 6 & 5 & 4 & 3 & 2 & 1 & 0 \\ \hline
	     	 B0 & 1.0 & 1.0 & 1.0 & 1.0 & 1.0 & 1.0 & 1.0 \\
	     	 B1 & 0.0 & 0.0 & 0.0 & 0.0 & 0.0 & 0.0 & 0.0 \\
	     	 B2 & 0.0 & 0.0 & 0.0 & 0.0 & 0.0 & 0.0 & 0.0 \\
	     	 B3 & 1.0 & 1.0 & 0.0 & 0.0 & 0.0 & 1.0 & 0.0 \\
	     	 B4 & 1.0 & 0.0 & 1.0 & 1.0 & 1.0 & 1.0 & 1.0 \\
	     	 B5 & 0.0 & 0.0 & 0.0 & 0.0 & 0.0 & 0.0 & 0.0	     	 
	     	\end{tabular}};	 		
	     		
	     \node[block, fill=gray!80, below of=f4, yshift=-3.25cm, text width=9cm] (fExitFuzzy) {
	     	\begin{tabular}{l|l|l|l|l|l|l|l|}
	     	 \textbf{Edge} & \multicolumn{7}{|c|}{\textbf{Insert}} \\
	     	 & 6 & 5 & 4 & 3 & 2 & 1 & 0 \\ \hline \\[-5pt]
	     	 B0$\rightarrow$B1 & 0.001 & 0.998 & 0.001 & 0.000 & 0.001 & 0.001 & 0.000 \\
	     	 	
	     	 B1$\rightarrow$B5 & 0.001 & 0.001 & 0.001 & 0.000 & 0.001 & 0.001 & 0.000 \\
	     	 
	     	 B1$\rightarrow$B2 & 0.001 & 0.001 & 0.001 & 0.000 & 0.001 & 0.001 & 0.000 \\
	     	 
	     	 B2$\rightarrow$B3 & 0.001 & 0.001 & 0.001 & 0.000 & 0.001 & 0.000 & 0.000 \\   
	     	 
	     	 B2$\rightarrow$B4 & 0.001 & 0.001 & 0.001 & 0.000 & 0.001 & 0.000 & 0.000 \\
	     	 
	     	 B3$\rightarrow$B4 & 0.000 & 0.998 & 0.001 & 0.000 & 0.001 & 0.000 & 0.000 \\
	     	 
	     	 B4$\rightarrow$B1 & 0.001 & 0.000 & 0.001 & 0.000 & 0.001 & 0.001 & 0.000 \\
			\\
	     	 \textbf{Block} & \multicolumn{7}{|c|}{\textbf{Delete}} \\
	     	 & 6 & 5 & 4 & 3 & 2 & 1 & 0 \\ \hline \\[-5pt]
	     	 \textbf{B0} & 0.000 & 0.000 & 0.000 & 0.000 & 0.000 & 0.000 & 0.000 \\
	     	 \textbf{B1} & 0.000 & 0.000 & 0.000 & 0.000 & 0.000 & 0.000 & 0.000 \\
	     	 \textbf{B2} & 0.000 & 0.000 & 0.000 & 0.000 & 0.000 & 0.001 & 0.000 \\
	     	 \textbf{B3} & 0.001 & 0.000 & 0.000 & 0.000 & 0.000 & 0.000 & 0.000 \\   
	     	 \textbf{B4} & 0.000 & 0.998 & 0.001 & 0.000 & 0.001 & 0.000 & 0.000 \\
	     	 \textbf{B5} & 0.000 & 0.000 & 0.000 & 0.000 & 0.000 & 0.000 & 0.000 
	     	\end{tabular}};
		 \draw[->, gray!20, line width=2pt] (fInitClas) -- (f1);
		 \draw[->, gray!80, line width=2pt] ($(fInitFuzzy.west)+(0,2)$) -- (f1.east);
		 			
		 \draw[->, gray!20, line width=2pt, bend right] (f1) edge (f2);	
		 \draw[->, gray!20, line width=2pt, bend right] (f2) edge (f3);
		 \draw[->, gray!20, line width=2pt, bend right] (f3) edge (f4);
		 
		 \draw[->, gray!80, line width=2pt, bend left] (f1) edge (f2);	
		 \draw[->, gray!80, line width=2pt, bend left] (f2) edge (f3);	 
		 \draw[->, gray!80, line width=2pt, bend left] (f3) edge (f4);	 
		 
		 \node[below right of=f4, xshift=4.5cm, yshift=0.7cm] (dummyNode0) {};
		 \node[below of=dummyNode0, yshift=-4.5cm] (dummyNode1) {};
		 \draw[gray!20, line width=2pt] (f4.east) -- (dummyNode0.center);
		 \draw[gray!20, line width=2pt] (dummyNode0.center) -- (dummyNode1.center);
		 \draw[->, gray!20, line width=2pt] (dummyNode1.center) -- (fExitClas.west);
		 \draw[->, gray!80, line width=2pt] (f4) -- (fExitFuzzy);
		 
	\end{tikzpicture}
	
	\vspace*{0.25cm}
	\hspace*{0.01cm}
	{\scriptsize
	 \begin{tabular}{|l||c|c|c|c|c|c|c|}
		 \hline
		 \textbf{Expression} & abs(a[i]-b) & Transform(b) & IncRate(i) & A*B & i+1 & in[i] != b & $i \leq N$ \\ \hline
		 \textbf{Index} & 6 & 5 & 4 & 3 & 2 & 1 & 0 \\
		 \hline
	 \end{tabular}	
	}
	\caption{Knoop-R\"{u}thing-Steffen LCM formulation (middle) using classical (left) and fuzzy (right/bottom) data-flow analysis}
	\label{fig:exLCMProb}
\end{figure}
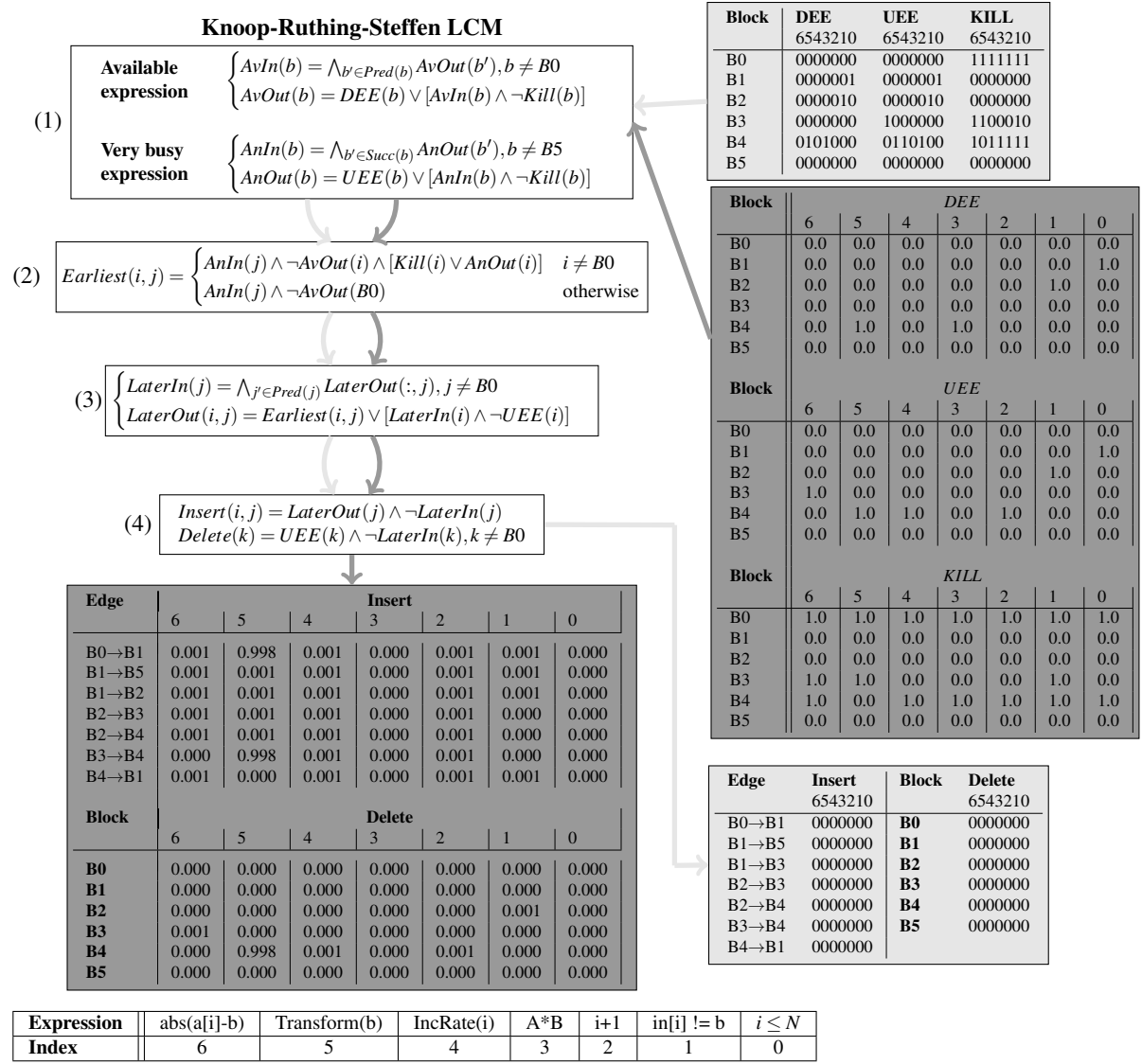

\noindent
Very busy expression analysis is a backward-must data-flow analysis that depends on $UEE$ and $KILL$ and computes the set of expressions that is guaranteed to be computed at some time in the future. Similarly Available expression analysis is a forward-must data-flow analysis that depends on $DEE$ and $KILL$ and deduces the set of previously computed expressions that may be reused. The fixed-point system of these two analyses are shown in Figure \ref{fig:exLCMProb}. It is beyond the scope of this paper to further elaborate on the details of these analyses, the interested reader should consider Nielson et al.~\cite{Nielson1999}. Here the LCM algorithm and the data-flow analyses it depends on, are applications we use to demonstrate the benefit of our framework. As such a rudimentary understanding is sufficient.

\newcommand{\colvecelem}[2]{$\left[\begin{tabular}{>{\hspace{-7pt}}c<{\hspace{-7pt}}}#1\\#2\end{tabular}\right]$}
\newcommand{\colvec}[5]{%
	\ifthenelse{\equal{#1}{0}}{\colvecelem{1}{0}}{\colvecelem{0}{1}}$\otimes$%
	\ifthenelse{\equal{#2}{0}}{\colvecelem{1}{0}}{\colvecelem{0}{1}}$\otimes$%
	\ifthenelse{\equal{#3}{0}}{\colvecelem{1}{0}}{\colvecelem{0}{1}}$\otimes$%
	\ifthenelse{\equal{#4}{0}}{\colvecelem{1}{0}}{\colvecelem{0}{1}}$\otimes$%
	\ifthenelse{\equal{#5}{0}}{\colvecelem{1}{0}}{\colvecelem{0}{1}}}
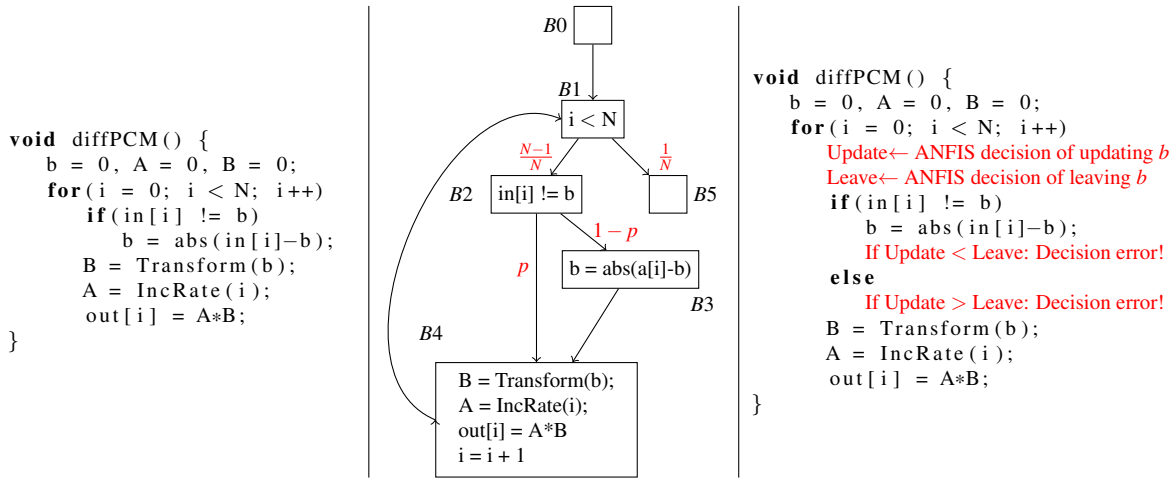
\begin{figure}
 \scriptsize
 \begin{minipage}{0.3\linewidth}
 	\begin{lstlisting}[language=C,mathescape,tabsize=3]
void diffPCM() {
	b = 0, A = 0, B = 0;
	for(i = 0; i < N; i++)
		if(in[i] != b)
		   b = abs(in[i]-b);
		B = Transform(b);
		A = IncRate(i);
		out[i] = A*B;
}
	\end{lstlisting}
 \end{minipage}%
 \vrule\hspace*{-0.75cm}
 \begin{minipage}{0.35\linewidth}
	\begin{tikzpicture}
			 \node[block] (f0) {}; 
			 \node[above of=f0,xshift=-0.5cm,yshift=-1cm] (f0Label) {$B0$};
			 
			 \node[block, below of=f0, yshift=-0.25cm] (f1) {i $<$ N};
			 \node[above of=f1,xshift=-0.3cm,yshift=-0.6cm] (f1Label) {$B1$};
			 
			 \node[block, below of=f1, xshift=-0.75cm] (f2) {in[i] != b};
			 \node[left of=f2] (f2Label) {$B2$};
			 
			 \node[block, below of=f2, xshift=1.25cm] (f3) {b = abs(a[i]-b)};
			 \node[below right of=f3,xshift=0.25cm,yshift=0.25cm] (f3Label) {$B3$};
			 
			 \node[block, below of=f3, xshift=-1.25cm, yshift=-1cm] (f4) {\begin{tabular}{l}B = Transform(b);\\ A = IncRate(i);\\ out[i] = A*B\\ i = i + 1\end{tabular}};		
			 \node[below of=f4,xshift=-1.4cm,yshift=2.15cm] (f4Label) {$B4$};
			 
			 \node[block, below of=f1, xshift=1cm] (f5) {};
			 \node[right of=f5, xshift=-0.5cm] (f5Label) {$B5$};	
			 \draw[->] (f0) -- (f1);
			 \draw[->] (f1) node[left,xshift=-0.4cm,yshift=-0.5cm,text=red]{$\frac{N-1}{N}$} -- (f2);
			 \draw[->] (f1) node[right,xshift=0.75cm,yshift=-0.5cm,text=red]{$\frac{1}{N}$} -- (f5);
			 \draw[->] (f2) node[right,xshift=0.65cm,yshift=-0.5cm,text=red]{$1-p$} -- (f3);
			 \draw[->] (f3) -- (f4);
			 \draw[->] (f2) node[left,yshift=-1cm,text=red]{$p$} -- (f4);
			 \path[draw,->,bend left=90] (f4.west) edge (f1.west);
	\end{tikzpicture}	
 \end{minipage}%
 \vrule\hspace*{0.1cm}
 \begin{minipage}{0.5\textwidth}
	\begin{lstlisting}[language=C,escapeinside={<@}{@>},mathescape,tabsize=3]
void diffPCM() {
	b = 0, A = 0, B = 0;
	for(i = 0; i < N; i++)
		<@\textcolor{red}{Update$\leftarrow$ ANFIS decision of updating $b$}@>
		<@\textcolor{red}{Leave$\leftarrow$ ANFIS decision of leaving $b$}@>
		if(in[i] != b)
		   b = abs(in[i]-b);
		   <@\textcolor{red}{If Update $<$ Leave: Decision error!}@>
		else
		   <@\textcolor{red}{If Update $>$ Leave: Decision error!}@>		
		B = Transform(b);
		A = IncRate(i);
		out[i] = A*B;
}
	\end{lstlisting}
 \end{minipage}
 \caption{\texttt{diffPCM} function (left), the corresponding flow-chart (middle) and the version used in Section \ref{sec:hybridanalysis} which is annotated with ANFIS classifier invocations (right)}
 \label{fig:diffPCM} 
\end{figure}

Consider the simplified differential pulse-code modulation routine \texttt{diffPCM} in Figure \ref{fig:diffPCM} (left). We assume that $N$ and the relative number of times block $B3$ (denoted $p$) is statically known\footnote{In this demonstration we let $p = 0.999$ and $N = 1000$, but our conclusions hold as $N$ increases and $p$ approaches $1$.}. In each iteration \texttt{diffPCM} invokes the pure functions \texttt{Transform}, to encode the differential output, and \texttt{IncRate} to get a quantification rate. We use the KRS-LCM algorithm to determine if these invocations can be made prior to entering the loop and contrast this to a situation where the data-flow analyses are performed in the fuzzy framework. As we will show the ``fuzzy KRS-LCM'' allows us to uncover opportunites the classical KRS-LCM would miss.

\subsection{Type-1 static analysis}
\label{sec:fLCMt1}
The data-flow problems of the KRS algorithm use expressions as domain. The mapping between expressions of \texttt{diffPCM} and indexes are listed in Figure \ref{fig:exLCMProb} (bottom) together with the values of $DEE$, $UEE$ and $KILL$ for each block (top right). The classical KRS algorithm conclude that both calls must be evaluated in B4 (bottom light gray box, ``Delete'' matrix, Column 4 and 5).

For the fuzzy data-flow analyses we use the Type-1 Min-Max fuzzy logic. The corresponding fuzzy sets of $DEE$, $UEE$ and $KILL$ are given in Figure \ref{fig:exLCMProb} (top dark gray box). Step (1) of the fuzzy KRS-LCM is hence the fixed-point to below system of equations:

{ \small
\begin{tabular}{ll}
 $\begin{array}{l} \textbf{Available expression} \\ \textbf{analysis system} \end{array}$ & $\begin{cases}
 	AvOut(B0) = 0.0 \\
 	AvOut(B1) = DEE(B1) \vee \left(\textcolor{red}{\left[\frac{1}{N}AvOut(B0) + \frac{N-1}{N}AvOut(B4)\right]} \wedge \lnot Kill(B1)\right) \\
 	AvOut(B2) = DEE(B2) \vee \left(\textcolor{red}{AvOut(B1)} \wedge \lnot Kill(B2)\right) \\
 	AvOut(B3) = DEE(B3) \vee \left(\textcolor{red}{AvOut(B2)} \wedge \lnot Kill(B3)\right) \\
 	AvOut(B4) = DEE(B4) \vee \left(\textcolor{red}{\left[pAvOut(B2) + (1-p)AvOut(B3)\right]} \wedge \lnot Kill(B4)\right) \\
 	AvOut(B5) = DEE(B5) \vee \left(\textcolor{red}{AvOut(B1)} \wedge \lnot Kill(B4)\right) \\
 \end{cases}$ \\
 \\
 $\begin{array}{l} \textbf{Very busy expression} \\ \textbf{analysis system} \end{array}$ & $\begin{cases}
 	AnOut(B0) = UEE(B0) \vee \left(\textcolor{red}{AnOut(B1)} \wedge \lnot Kill(B1)\right) \\
 	AnOut(B1) = UEE(B1) \vee \left(\textcolor{red}{\left[\frac{N-1}{N}AnOut(B2) + \frac{1}{N}AnOut(B5)\right]} \wedge \lnot Kill(B1)\right) \\
 	AnOut(B2) = UEE(B2) \vee \left(\textcolor{red}{\left[pAnOut(B4) + (1-p)AnOut(B3)\right]} \wedge \lnot Kill(B2)\right) \\
 	AnOut(B3) = UEE(B3) \vee \left(\textcolor{red}{AnOut(B4)} \wedge \lnot Kill(B3)\right) \\
 	AnOut(B4) = UEE(B4) \vee \left(\textcolor{red}{AnOut(B1)} \wedge \lnot Kill(B4)\right)\\
 	AnOut(B5) = 0.0 \\
 \end{cases}$
\end{tabular}
}

Steps (2) and (4) introduce (constant) predicates and are performed outside the analysis framework. Step (3) is done similarly to step (1). 
Figure \ref{fig:exLCMProb} (bottom dark gray box) shows the result from step (4). In contrast to the classical LCM the result implies that it is \textit{very plausible} (0.998) that we can delete the invocation of \texttt{Transform} (``Delete'' matrix, Column 5) from block B4 and instead add it at the end of B0 and B3 (or start of B1 and B4). However, result for the invocation of \texttt{IncRate} remains. This is because the invocation depends on the value of $i$ which is updated at the end of B4.

\subsection{Type-2 static analysis}
To increase data-flow analysis precision a function call is sometimes inlined at the call site.
The improvement can however be reduced if the control-flow analysis is inaccurate and multiple targets are considered for a particular call site. We show how the uncertainty in control-flow and data-flow can be quantified in two different dimensions using type-2 interval fuzzy sets. As per Section \ref{sec:prelim} we can lift an arbitrary fuzzy predicate to intervals. Here we assume no knowledge about the relative number of calls to each target and treat the different calls non-deterministically.

\begin{figure}[!ht]
	{ \linespread{0.9}\fontfamily{ptm}
	\begin{minipage}{0.5\textwidth}
	  \begin{tikzpicture}
		 \node[block, fill=gray!10] (f4) {\begin{tabular}{l}B = Transform(b);\\ ... = IncRate(i);\\ \\ \\ \\ \\ \\ \\ A = ...\\ out[i] = A*B\\ i = i + 1\end{tabular}};		
		 \node[block, fill=white, xshift=-2cm] (fIncRateA) {%
\begin{lstlisting}[language=C,mathescape,basicstyle=\tiny,tabsize=3]
int IncRate_1(int i) {
	return 2*i;
}
\end{lstlisting}};
		 \node[block, fill=white, xshift=2cm] (fIncRateB) {%
\begin{lstlisting}[language=C,mathescape,basicstyle=\tiny,tabsize=3]
int IncRate_2(int i) {
	return 1;
}
\end{lstlisting}};
		 \draw[->] ($(f4)+(-0.15,1.0)$) -- (fIncRateA);
		 \draw[->] ($(f4)+(-0.15,1.0)$) -- (fIncRateB);		
		 \draw[->] (fIncRateA) -- ($(f4)+(-0.15,-1.0)$);
		 \draw[->] (fIncRateB) -- ($(f4)+(-0.05,-1.0)$);
	  \end{tikzpicture}	
	\end{minipage}%
	\begin{minipage}{0.45\textwidth}
		\scriptsize
		\begin{tabular}{|l|lll|}
		  \hline
		  \textbf{Block} & \textbf{Kill} & \textbf{DEE} & \textbf{UEE} \\ \hline
		  6 & $\left[1.0, 1.0\right]$ & $\left[0.0, 0.0\right]$ & $\left[0.0, 0.0\right]$ \\
		  5 & $\left[0.0, 0.0\right]$ & $\left[1.0, 1.0\right]$ & $\left[1.0, 1.0\right]$ \\
		  4 & $\left[1.0, 1.0\right]$ & $\left[0.0, 1.0\right]$ & $\left[1.0, 1.0\right]$ \\
		  3 & $\left[1.0, 1.0\right]$ & $\left[1.0, 1.0\right]$ & $\left[0.0, 0.0\right]$ \\
		  2 & $\left[1.0, 1.0\right]$ & $\left[0.0, 0.0\right]$ & $\left[1.0, 1.0\right]$ \\
		  1 & $\left[1.0, 1.0\right]$ & $\left[0.0, 0.0\right]$ & $\left[0.0, 0.0\right]$ \\
		  0 & $\left[1.0, 1.0\right]$ & $\left[0.0, 0.0\right]$ & $\left[0.0, 0.0\right]$ \\ \hline
		\end{tabular}\\
		\begin{tabular}{|l|l||l|l|}
			\hline
	    	\textbf{Edge} & \textbf{Insert} & \textbf{Block} & \textbf{Delete} \\ \hline
			B0$\rightarrow$B1 & $\left[0.001, 0.999\right]$ & %
				B0 & $\left[0.000, 0.000\right]$ \\
			B1$\rightarrow$B5 & $\left[0.001, 0.999\right]$ & %
				B1 & $\left[0.000, 0.000\right]$ \\
			B1$\rightarrow$B3 & $\left[0.001, 0.999\right]$ & %
				B2 & $\left[0.000, 0.000\right]$ \\
			B2$\rightarrow$B3 & $\left[0.001, 0.999\right]$ & %
				B3 & $\left[0.000, 0.000\right]$ \\
			B2$\rightarrow$B4 & $\left[0.001, 0.999\right]$ & %
				B4 & $\left[0.002, 0.999\right]$ \\
			B3$\rightarrow$B4 & $\left[0.001, 0.999\right]$ & %
				B5 & $\left[0.000, 0.000\right]$ \\
			B4$\rightarrow$B1 & $\left[0.000, 0.999\right]$ & & \\
			\hline
		\end{tabular}
	\end{minipage}
	\caption{Implementations of \texttt{IncRate} inlined in block B4 (left); DEE, UEE and Kill vectors of block B4 and \textit{Delete}  \textit{Insert} analysis result for expression \texttt{IncRate(i)} (right)}
	\label{fig:interproc}
	}	
\end{figure}
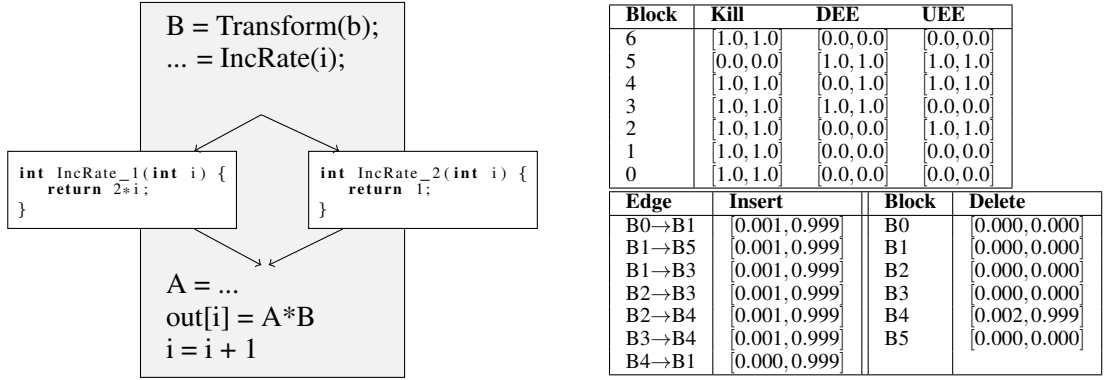
We assume two different \texttt{IncRate} functions, as in Figure \ref{fig:interproc} (left), have been determined as targets. Their respective $UEE$ and $Kill$ entries are the same but since $i$ is updated at the end of block B4 their $DEE$ entry will differ. The result of \texttt{IncRate\_1} depends on the variable $i$ and therefore $DEE(B4_1) = \mathbf{0101000}$, in contrast the entry for \texttt{IncRate\_2} is $DEE(B4_2) = \mathbf{0111000}$, where $\mathbf{0} = [0,0]$ and $\mathbf{1} = [1,1]$. The new entry for block B4 is given by $DEE(B4) = DEE(B4_1) \fuzzyvee DEE(B4_2) = \left<\mathbf{0},\mathbf{1},[0,1],\mathbf{1},\mathbf{0},\mathbf{0},\mathbf{0}\right>$. The new $Kill$, $DEE$ and $UEE$ sets are given in Figure \ref{fig:interproc} (right). 

Applying the fuzzy KRS-LCM, but with Type-1 min-max fuzzy logic lifted to Interval type-2 min-max fuzzy logic gives the values of \textit{Delete} and \textit{Insert} for expression \texttt{IncRate(i)} in Figure \ref{fig:interproc} (right). The result for invoking \texttt{IncRate} prior to the loop is $[0.001, 0.999]$ as opposed to 0.001 from the Type-1 analysis in Section \ref{sec:fLCMt1}. The added dimension in the result of the type-2 fuzzy analysis allows us to differentiate uncertain results from pessimistic results. In the given example we showed that the result of Section \ref{sec:fLCMt1} is a pessimistic over-generalization and that the two paths need to be considered seperately to increase precision.

\subsection{Hybrid analysis}
\label{sec:hybridanalysis}
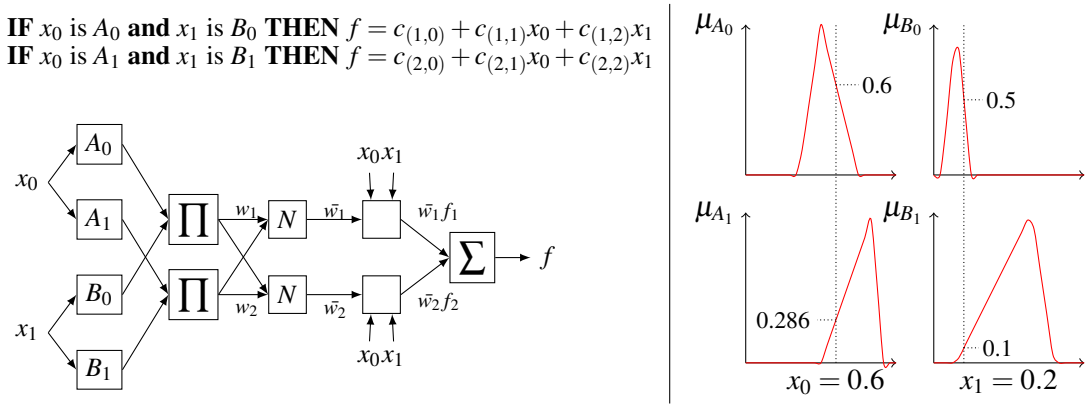
\begin{figure}
 \begin{minipage}{0.55\linewidth}
 	\linespread{0.9}\fontfamily{ptm}\selectfont\small
  	\textbf{IF} $x_0$ is $A_0$ \textbf{and} $x_1$ is $B_0$ \textbf{THEN} $f = c_{(1,0)} + c_{(1,1)}x_0 + c_{(1,2)}x_1$\newline
  	\textbf{IF} $x_0$ is $A_1$ \textbf{and} $x_1$ is $B_1$ \textbf{THEN} $f = c_{(2,0)} + c_{(2,1)}x_0 + c_{(2,2)}x_1$\newline
 	\newline
 	\newline
	\tikzset{>=latex}
 	\begin{tikzpicture}
	    \node (x0) {$x_0$};
	    \node[below of=x0, yshift=-1cm] (x1) {$x_1$};
	    \node[block, above right of=x0, xshift=0.25cm, yshift=-0.20cm] (A0) {$A_0$};
	    \node[block, below of=A0] (A1) {$A_1$};
	    \node[block, below of=A1] (B0) {$B_0$};
	    \node[block, below of=B0] (B1) {$B_1$};
	    \node[block, right of=A1, xshift=0.25cm] (P0) {$\mathlarger{\prod}$};
	    \node[block, below of=P0] (P1) {$\mathlarger{\prod}$};
	    \node[block, right of=P0, xshift=0.25cm] (N0) {$N$};
	    \node[block, below of=N0] (N1) {$N$};
	    \node[block, right of=N0, xshift=0.25cm] (F0) {};
	     \node[above of=F0, yshift=-0.15cm, xshift=-0.16cm] (F0x0) {$x_0$};
	     \node[above of=F0, yshift=-0.15cm, xshift=0.16cm] (F0x1) {$x_1$};
     
	     \draw[->] (F0x0.south) -- ([xshift=-4]F0.north);
	     \draw[->] (F0x1.south) -- ([xshift=4]F0.north);
	    \node[block, below of=F0] (F1) {};
	     \node[below of=F1, yshift=0.15cm,xshift=-0.16cm] (F1x0) {$x_0$};
	     \node[below of=F1, yshift=0.15cm,xshift=0.16cm] (F1x1) {$x_1$};
     
	     \draw[->] (F1x0.north) -- ([xshift=-4]F1.south);
	     \draw[->] (F1x1.north) -- ([xshift=4]F1.south);
	    \node[block, below right of=F0, yshift=0.2cm,xshift=0.5cm] (AVG) {$\mathlarger{\sum}$};
	    \node[right of=AVG] (O) {$f$};
    
	   \draw[->] (x0.east) -- (A0.west);
	   \draw[->] (x0.east) -- (A1.west);
	   \draw[->] (x1.east) -- (B0.west);
	   \draw[->] (x1.east) -- (B1.west);
   
	   \draw[->] (A0.east) -- (P0.west);
	   \draw[->] (B0.east) -- (P0.west);
	   \draw[->] (A1.east) -- (P1.west);
	   \draw[->] (B1.east) -- (P1.west);
   
	   \draw[->] (P0.east) node[right,yshift=0.1cm,xshift=0.1cm]{\scriptsize $w_1$} -- (N0.west);
	   \draw[->] (P0.east) -- (N1.west);
	   \draw[->] (P1.east) node[right,yshift=-0.2cm, xshift=0.1cm]{\scriptsize $w_2$} -- (N1.west);
	   \draw[->] (P1.east) -- (N0.west);
   
	   \draw[->] (N0.east) node[right,yshift=0.1cm, xshift=0.1cm]{\scriptsize $\bar{w_1}$} -- (F0.west);
	   \draw[->] (N1.east) node[right,yshift=-0.2cm, xshift=0.1cm]{\scriptsize $\bar{w_2}$} -- (F1.west);
   
	   \draw[->] (F0.east) node[right,yshift=0.1cm,xshift=0.1cm]{\scriptsize $\bar{w_1}f_1$} -- (AVG.west);
	   \draw[->] (F1.east) node[right,yshift=-0.1cm,xshift=0.1cm]{\scriptsize $\bar{w_2}f_2$} -- (AVG.west);
   
	   \draw[->] (AVG.east) -- (O.west);
 	\end{tikzpicture}
 \end{minipage}%
 \vrule\hspace*{0.1cm}
 \begin{minipage}{0.4\linewidth}
  \begin{tikzpicture}[domain=0:1]
    \draw[->] (0,0) -- (2,0);
    \draw[->] (0,0) -- (0,2) node[left] {$\mu_{A_0}$};
    \draw[color=red,scale=2,smooth,variable=\x] plot[id=Rule1A0] function{(x <= 0.35 ? 0 : (x <= 0.5 ? (x-0.35)/(0.5-0.35) : (x <= 0.75 ? (0.75-x)/(0.75-0.5) : 0)))} (0,0);
    
    \draw[->] (2.5,0) -- (4.5,0);
    \draw[->] (2.5,0) -- (2.5,2) node[left] {$\mu_{B_0}$};
    \draw[color=red,scale=2,smooth,xshift=1.25cm,variable=\x] plot[id=Rule1B0] function{(x <= 0.05 ? 0 : (x <= 0.15 ? (x-0.05)/(0.15-0.05) : (x <= 0.25 ? (0.25-x)/(0.25-0.15) : 0)))};
    
    \draw[->] (0,-2.5) -- (2,-2.5);
    \draw[->] (0,-2.5) -- (0,-0.5) node[left] {$\mu_{A_1}$};
    \draw[color=red,scale=2,smooth,yshift=-1.25cm,variable=\x] plot[id=Rule2A1] function{(x <= 0.5 ? 0 : (x <= 0.85 ? (x-0.5)/(0.85-0.5) : (x <= 0.9 ? (0.9-x)/(0.9-0.85) : 0)))};
    
    \draw[->] (2.5,-2.5) -- (4.5,-2.5);
    \draw[->] (2.5,-2.5) -- (2.5,-0.5) node[left] {$\mu_{B_1}$};
    \draw[color=red,scale=2,smooth,xshift=1.25cm,yshift=-1.25cm,variable=\x] plot[id=Rule2B1] function{(x <= 0.15 ? 0 : (x <= 0.65 ? (x-0.15)/(0.65-0.15) : (x <= 0.8 ? (0.8-x)/(0.8-0.65) : 0)))};
    
	    \draw[densely dotted] (1.2,-2.5) -- (1.2,2); 
	    \draw[densely dotted] (1.2,1.2) -- (1.5,1.2); 
	    \node (x0A0) at (1.75,1.2) {\smaller 0.6};

	    \draw[densely dotted] (2.9,-2.5) -- (2.9,2); 
	    \draw[densely dotted] (2.9,1) -- (3.15,1); 
	    \node (x1B0) at (3.4,1) {\smaller 0.5};
	    \node (x0A1) at (1.2,-2.75) {$x_0 = 0.6$}; 
	    \draw[densely dotted] (1.2,-1.928571429) -- (0.95,-1.928571429); 
    	\node (x0A1) at (0.5,-1.928571429) {\smaller 0.286};
	    
    	\node (x1B1) at (3.5,-2.75) {$x_1 = 0.2$};
		\draw[densely dotted] (2.9,-2.3) -- (3.1,-2.3); 
		\node (x1B1) at (3.35,-2.3) {\smaller 0.1};
  \end{tikzpicture}

 \end{minipage} 
 \caption{First-order Takagi-Sugeno ANFIS with two rules and two variables (left) and four example fuzzy sets (right)}
 \label{fig:tsanfis}
\end{figure}

The result from a fuzzy data-flow analysis is a set of fuzzy membership degrees.
This section shows how the result can automatically be improved following the static analysis using a fuzzy regulator/classifier, if more specific information is provided at a later point. The classifier, a Takagi-Sugeno Adaptive-Network-based fuzzy inference system (TS-ANFIS)~\cite{jang93,jang96} shown in Figure \ref{fig:tsanfis}, is composed of five layers:
\begin{enumerate}[noitemsep,topsep=0pt,parsep=0pt,partopsep=0pt]
 \item Lookup fuzzy membership degree of the input value.
 \item Compute the \textit{firing strength of a rule}, i.e. conjunction of all membership degrees from each rule.
 \item Normalize the firing strengths, i.e., $\bar{w_i} = w_i/\sum_j w_j$.
 \item Weight the normalized firing strength to the consequent output of the rule $f_i(x)$.
 \item Combine all rule classifiers, i.e. $f=\sum_i \bar{w_i}f_i$ .
\end{enumerate}

\noindent
This classifier uses a polynomial (i.e., the consequent part of the adaptive IF-THEN rules) to decide the output membership. The order of the TS-ANFIS is the order of the polynomial. The classification accuracy of the TS-ANFIS can be improved online/offline by fitting the polynomial to the input data. For a first-order TS-ANFIS this can be implemented as follows:
\begin{itemize}
 \item (\textit{Offline}) (Affine) Least square (LS) optimization \cite{jang93} is a convex optimization problem that finds an affine function (i.e., $y = a_0 + \sum_{i = 1}^n a_ix_i$) which minimizes $||A[1; X] - Y||^2_2$ where $X$ and $Y$ are the input and output vectors of the training set.
 \item (\textit{Online}) Least mean square (LMS) \cite{jang93} is an adaptive filter that gradually (in steps of a given constant $\mu$) minimizes $\mathbb{E}\left[|y - f(x)|^2\right]$, where $\left<x,y\right>$ is an input/output sample.
\end{itemize}
To exemplify the functionality of the TS-ANFIS we consider the classification of $\vec{x} = \left<0.6,0.2\right>$ using the two rule TS-ANFIS from Figure \ref{fig:tsanfis} (left). Let $f_1(\vec{x}) = 0.2x_0 - 0.43x_1$, $f_2(\vec{x}) = 0.1x_1 + 0.5$ and membership functions be given as in Figure \ref{fig:tsanfis} (right). The membership degrees are marked in the figure as $\mu_{A0}(x0) = 0.6$, $\mu_{B0}(x1) = 0.5$ for the first rule and $\mu_{A1}(x0) = 0.286$, $\mu_{B1}(x0) = 0.1$ for the second rule. Hence the weight of the first rule (i.e., $w_1$) is $0.6 \wedge 0.5 = 0.5$ and the second rule (i.e., $w_2$) is $0.286 \wedge 0.1 = 0.1$. The normalized weights are then $\bar{w_1} = 0.833$ and $\bar{w_1} = 0.167$. As the consequence functions output $f_1(\vec{x}) = 0.034$ and $f_2(\vec{x}) = 0.52$ we produce the prediction $0.833f_1(\vec{x}) + 0.167f_2(\vec{x}) = 0.115$.  	

We return to the \texttt{diffPCM} function and again consider if we can invoke \texttt{Transform(b)} prior to entering the loop. We saw in Section \ref{sec:fLCMt1} that the fuzzy membership degree was 0.998. To improve classification accuracy we let the TS-ANFIS also use the $i$ variable and the first input value (i.e., $in[0]$). These variables were not part of the analysis and so we conservatively assume the fuzzy membership degree to be the same for any value of these variables (in our experiments: $1.0$). As shown in Figure \ref{fig:diffPCM} (right), we inserted calls to compute the ANFIS decision of updating and keeping the variable $b$ constant in the \texttt{diffPCM} function. If the incorrect decision was made the error was noted and an error rate computed after handling all input samples. 

\begin{figure}
	\centering
    \begin{subfigure}[t]{0.4\textwidth}
        \centering
        \hspace*{-1.5cm}
        \includegraphics[height=2in]{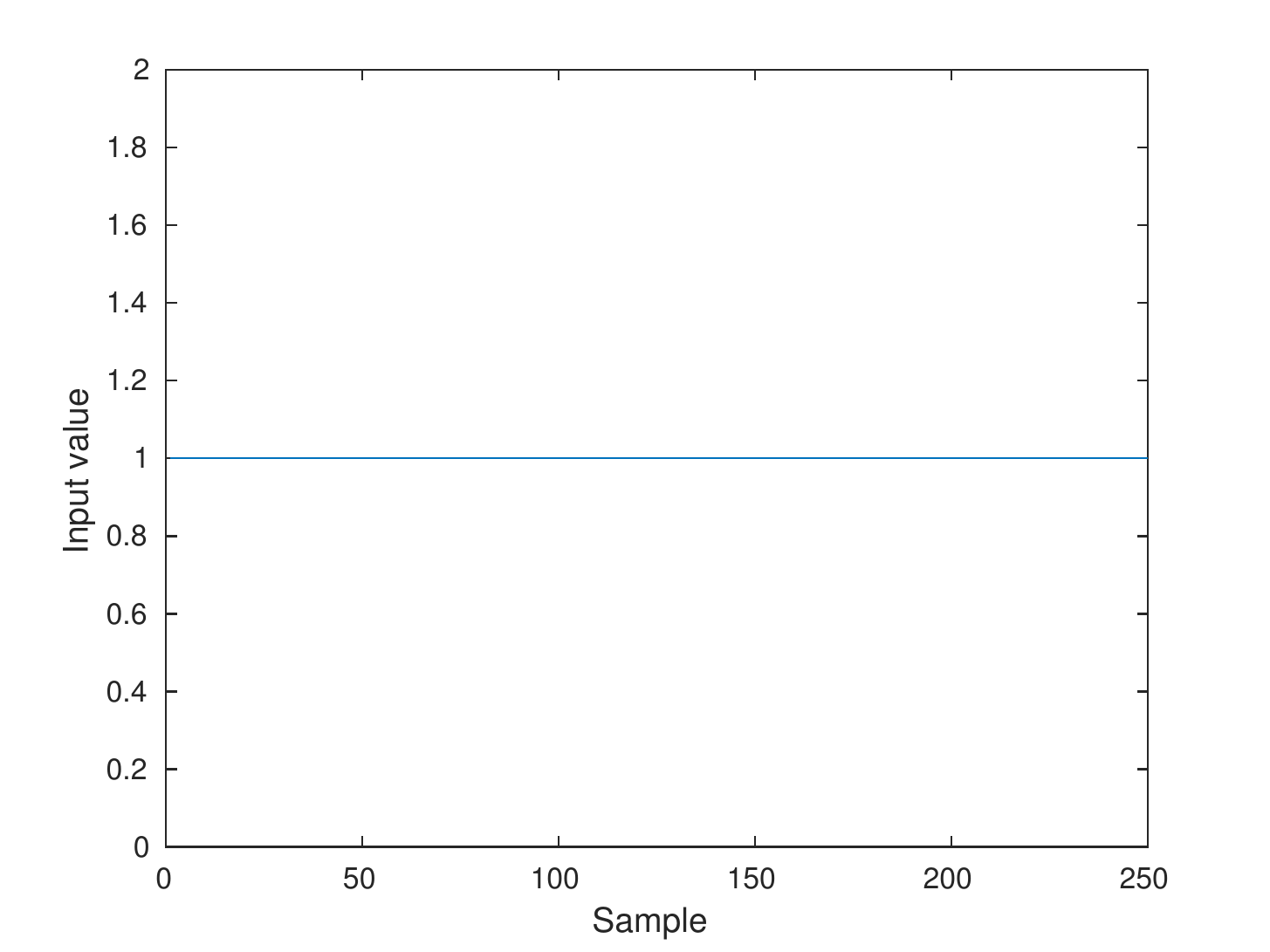}
    \end{subfigure}%
    ~ 
    \begin{subfigure}[t]{0.4\textwidth}
        \centering
        \includegraphics[height=2in]{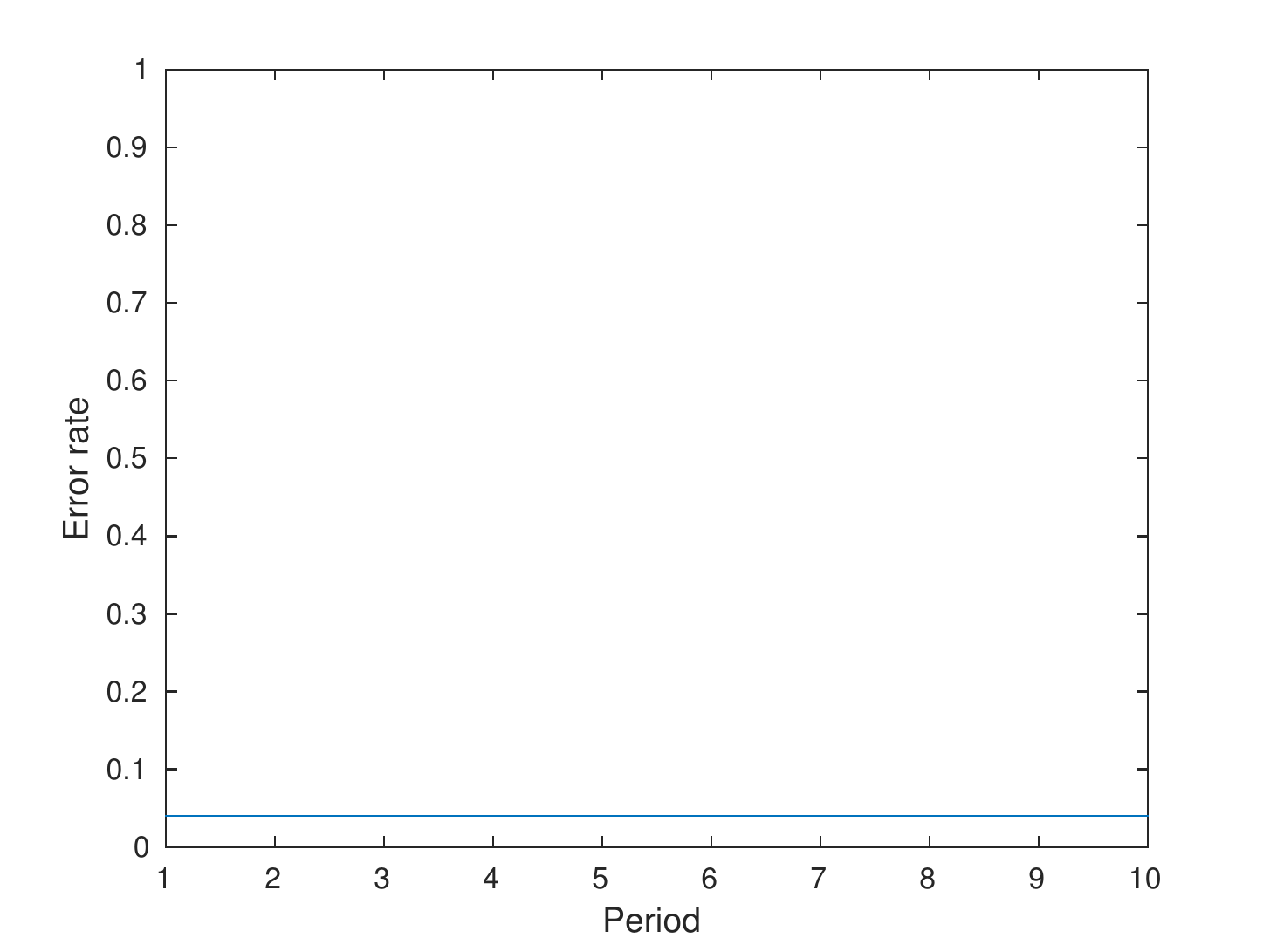}
    \end{subfigure}
    
    \begin{subfigure}[t]{0.4\textwidth}
        \centering
        \hspace*{-1.5cm}
        \includegraphics[height=2in]{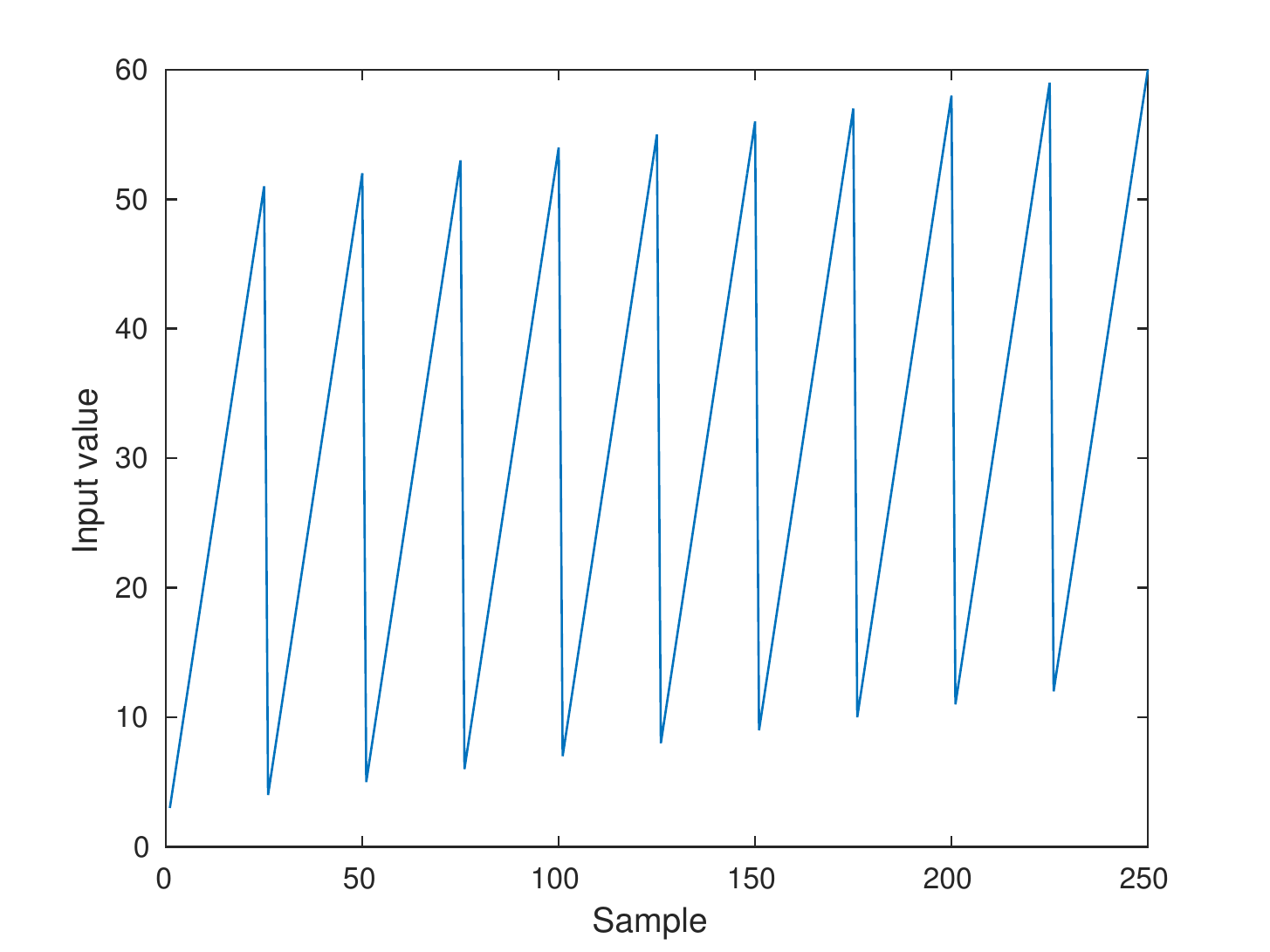}
    \end{subfigure}%
    ~ 
    \begin{subfigure}[t]{0.4\textwidth}
        \centering
        \includegraphics[height=2in]{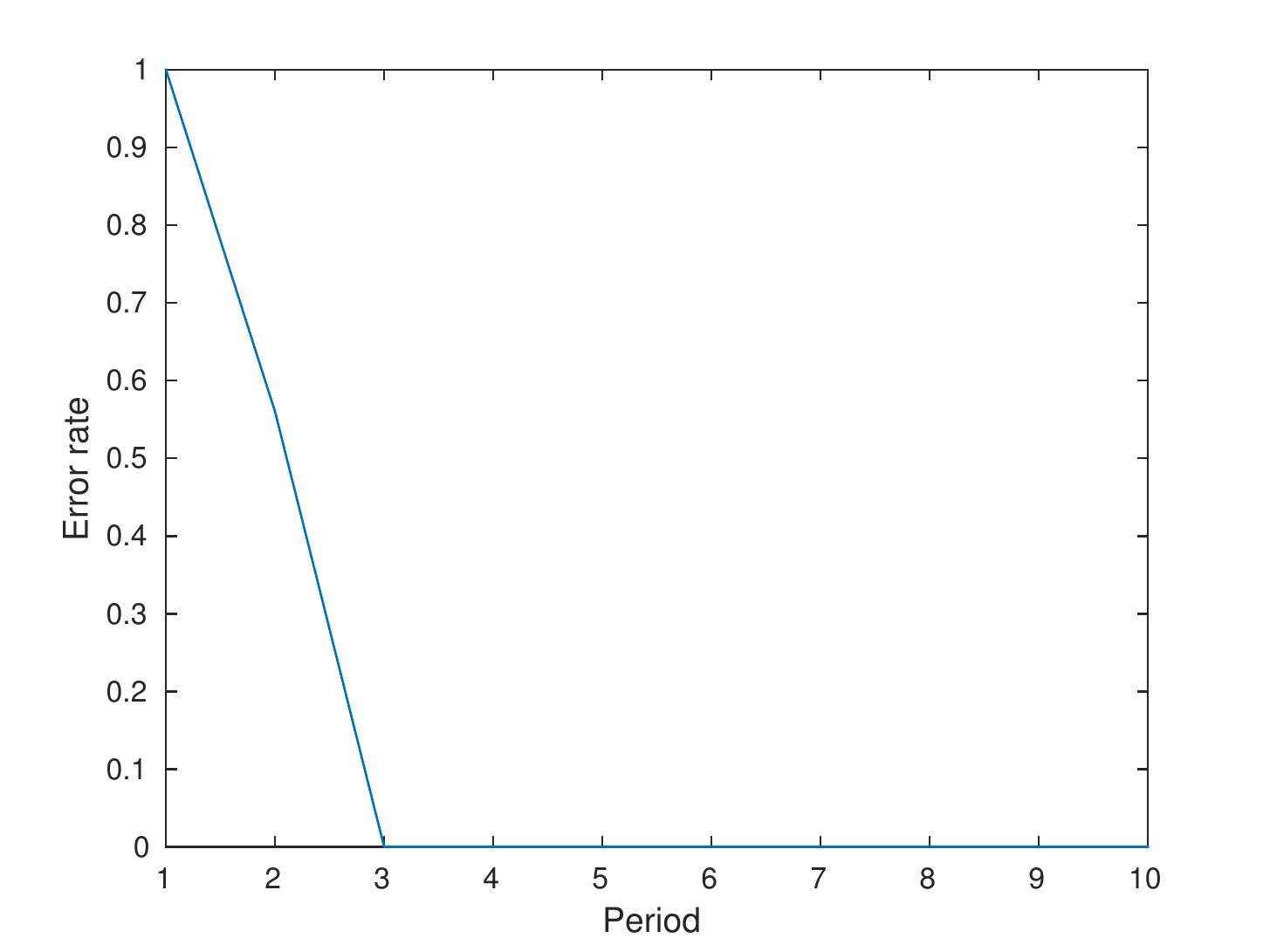}
    \end{subfigure}
    
    \begin{subfigure}[t]{0.4\textwidth}
        \centering
        \hspace*{-1.5cm}
        \includegraphics[height=2in]{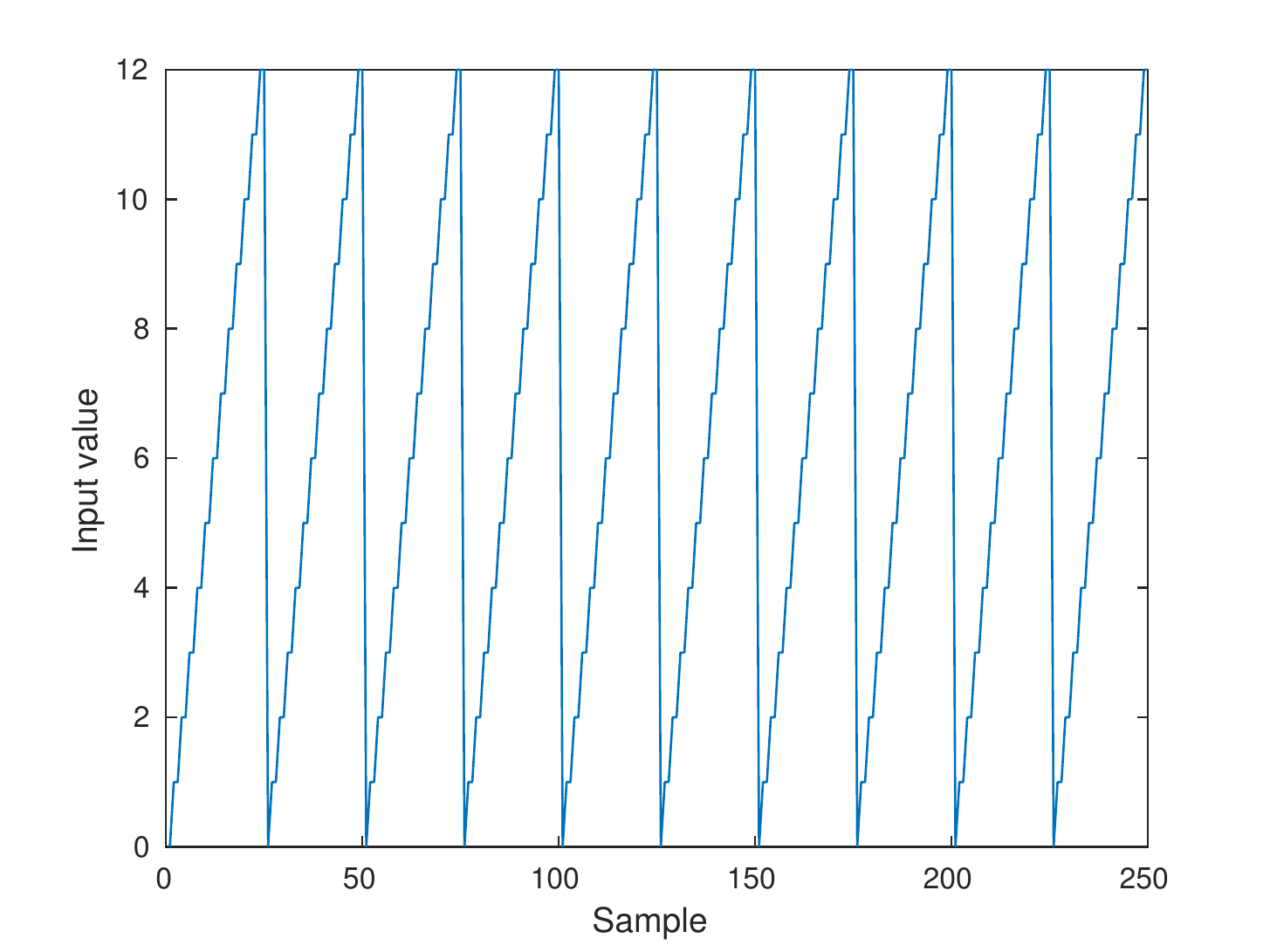}
    \end{subfigure}%
    ~ 
    \begin{subfigure}[t]{0.4\textwidth}
        \centering
        \includegraphics[height=2in]{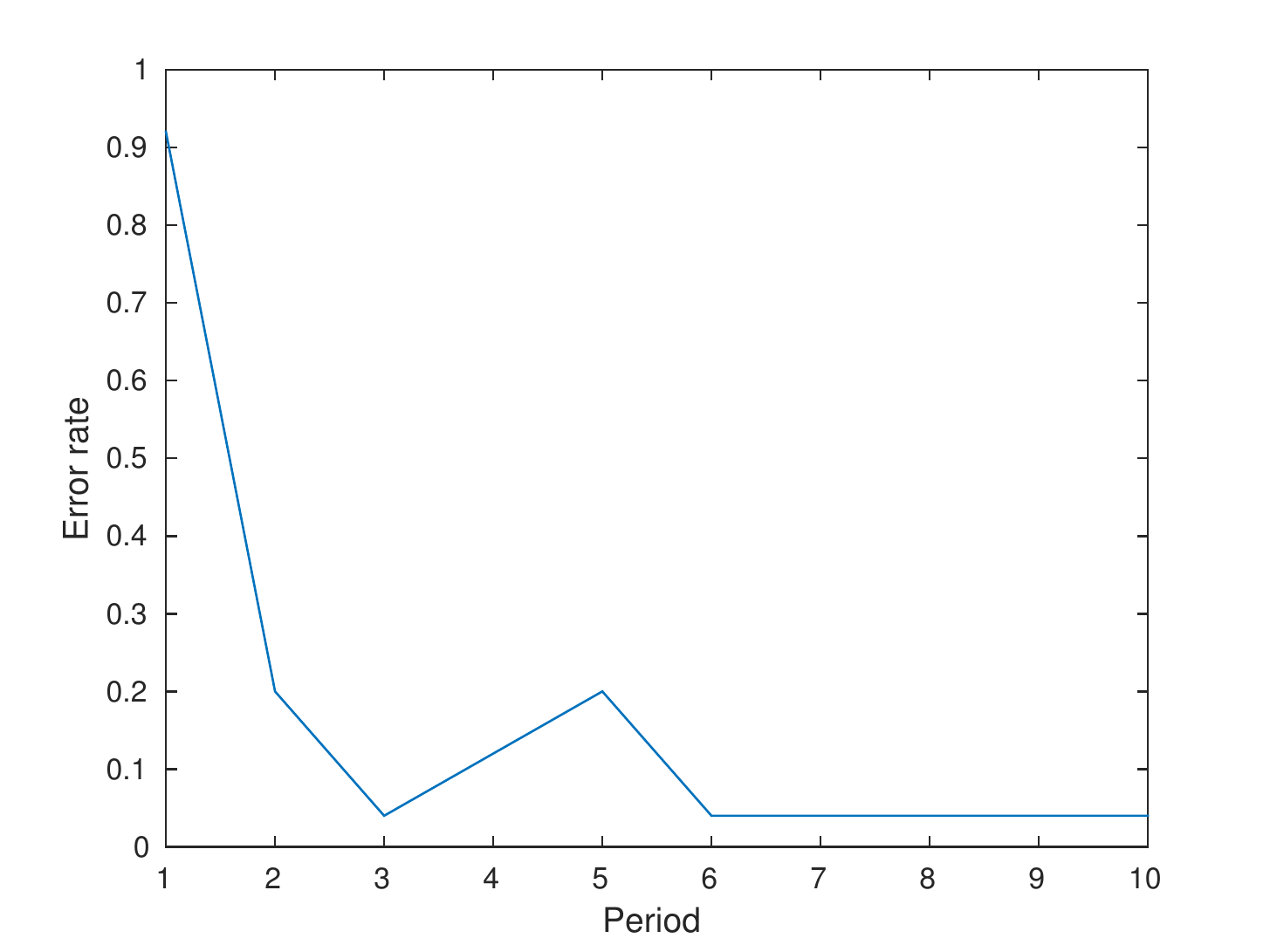}
    \end{subfigure}
    
    \begin{subfigure}[t]{0.4\textwidth}
        \centering
        \hspace*{-1.5cm}
        \includegraphics[height=2in]{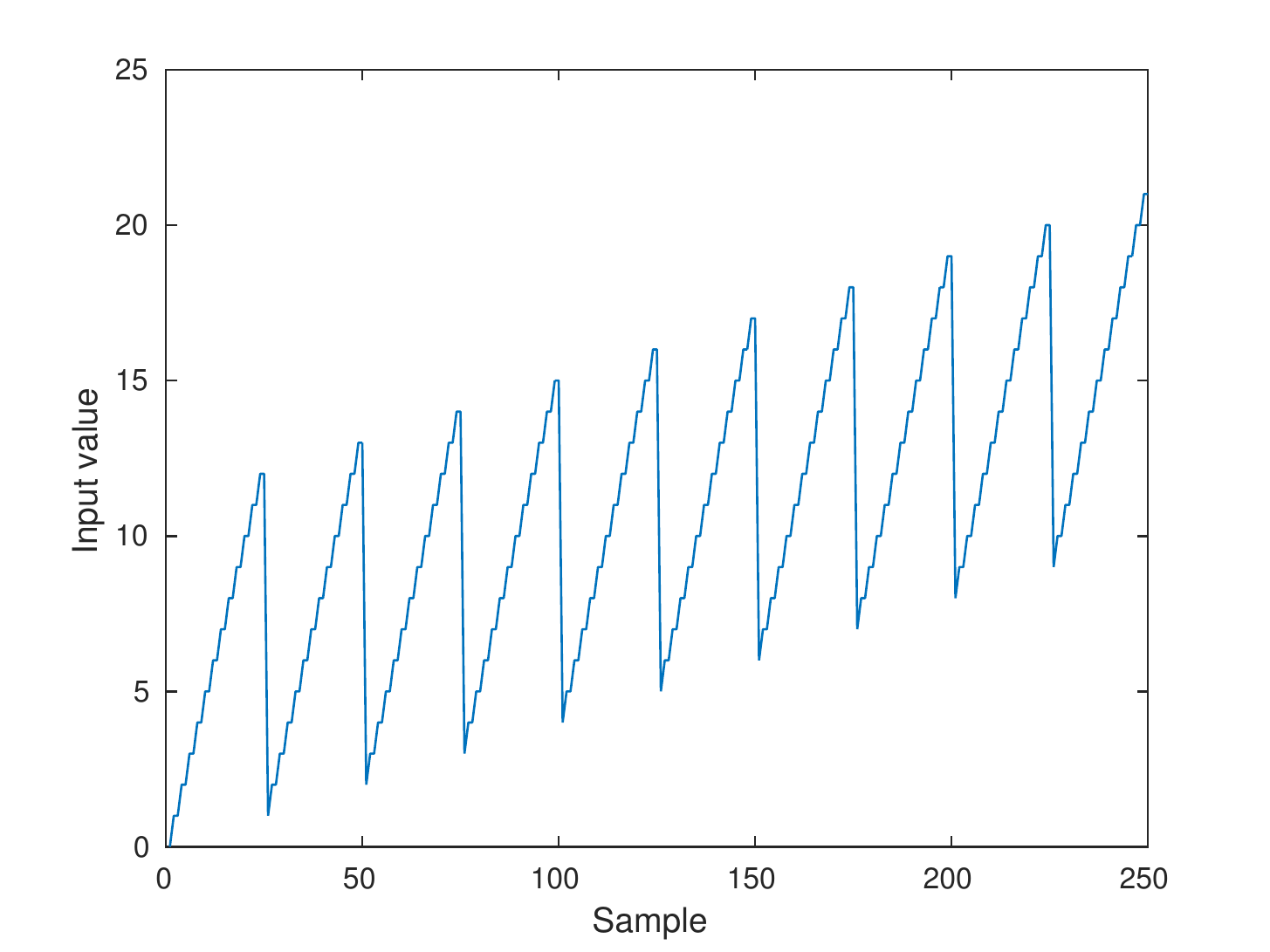}
    \end{subfigure}%
    ~ 
    \begin{subfigure}[t]{0.4\textwidth}
        \centering
        \includegraphics[height=2in]{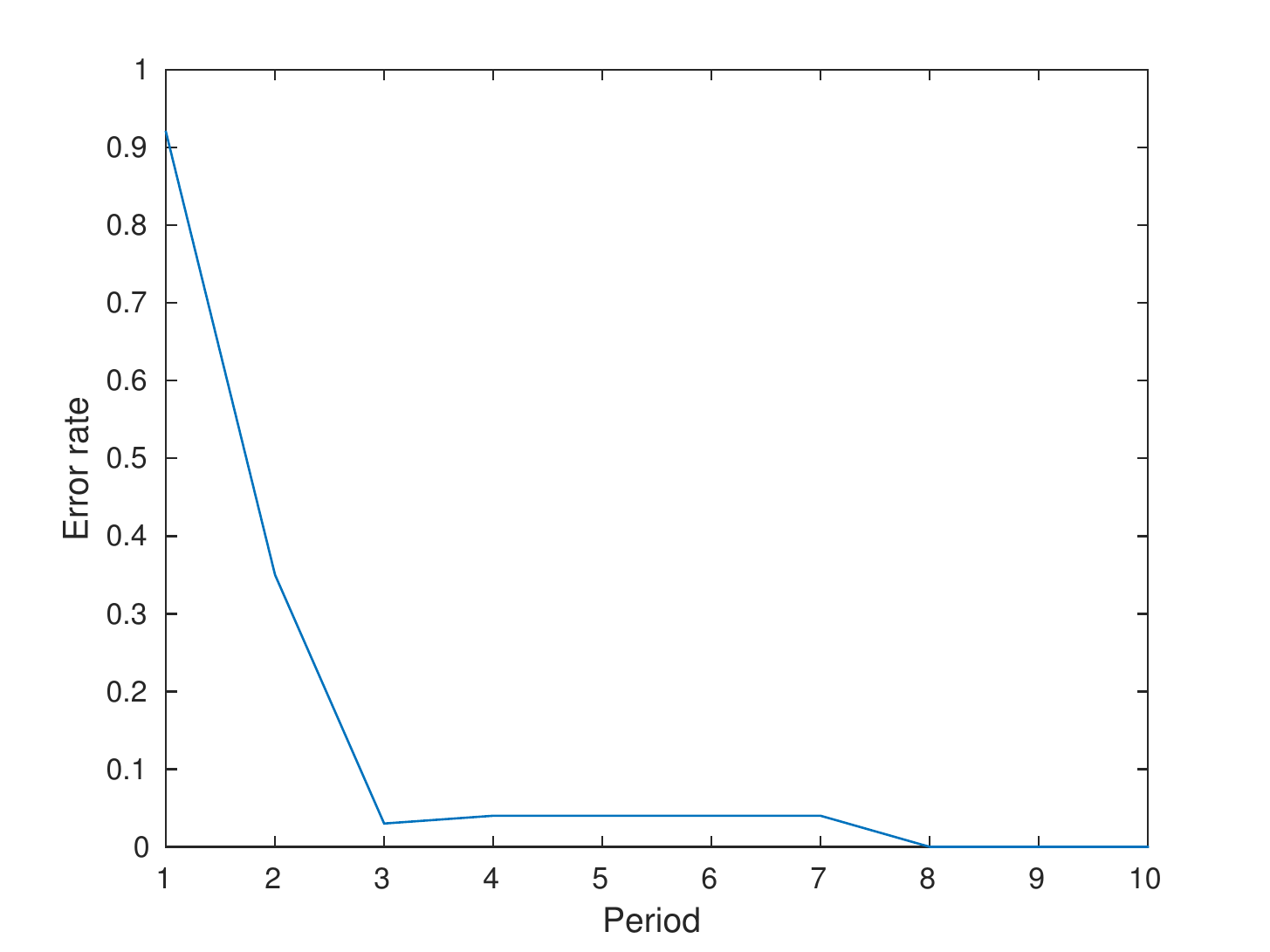}
    \end{subfigure}  
     
    \caption{$10\times 25$ input values (left) and the corresponding classification error rate (right)}
    \label{fig:anfis}
\end{figure}

We consider invoking the \texttt{diffPCM} function on four different input sets. Each input set defined as 10 periods with 25 input values in each period. The input sets (i.e., \texttt{in[...]}) is given in Figure \ref{fig:anfis} (left). We use the LMS algorithm\footnote{The constant $\mu$ for the four different runs was set to 0.001, 0.05, 0.15 and 0.1 respectively.} after each incorrect classification and the LS algorithm if the error rate of a period was larger than or equal to $80\%$. Note that the values of a period is not always perfectly representable by a linear classifier and sometimes varies between different periods, although periods are ``\textit{similar}''. Hence we do not expect the classifier to be monotonically improving with increasing period.  
As shown in the result in Figure \ref{fig:anfis} (right) the classification error decreases fast with both period and input sample. In two cases a small residual error remains after the final period. This show that the TS-ANFIS can improve the analysis result dynamically and hence increase the accuracy of when \texttt{Transform} can be invoked prior to entering the loop.

\section{Related work}
\label{sec:related}
Most systems include elements (e.g., input values, environment state) where information is limited but probabilistic and/or non-deterministic uncertainty can be formulated. For these systems a \textit{most likely} or even \textit{quantitative analysis} of properties is possible. Often this analysis relies on a probability theory for logical soundness. Cousot and Monerau~\cite{CousotMonerau-ESOP2012-PAI} introduced a unifying framework for probabilistic abstract interpretation. Much work have since, although perhaps implicitly, relied on their formulation. Often probabilistic descriptions are known with imprecision that manifests as non-deterministic uncertainty~\cite{Cousot2014}. Adje et al.~\cite{Assale14} introduced an abstraction based on the zonotope abstraction for Dempster-Shafer structures and P-boxes\footnote{Lower and upper bounds on a cumulative probability distribution functions}.

Di Pierro et al.~\cite{diPierro07} developed a probabilistic abstract interpretation framework and demonstrated an alias analysis algorithm that could guide the compiler in this decision. They later formulated data-flow problems (e.g., liveness analysis) in the same framework \cite{DiPierro13}. An important distinction between their (or similar probabilistic frameworks) and classical frameworks is the definition of the confluence operator. In contrast to a classical may- or must framework they use the weighted average. This is similar to the work by Ramalingam~\cite{Ramalingam:1996} that showed that the meet-over-paths (MOP) solution exists for such confluence operator with a transfer function defined in terms of min, max and negation (i.e., the Min-max fuzzy logic). Our work extends this to allow other transfer functions and integrates the static data-flow analysis with a dynamic refinement mechanism through fuzzy control theory.

\section{Conclusion}
A major problem for static program analysis is the limited input information and hence the conservative results. To alleviate the situation dynamic program analysis is sometimes used. Here accurate information is available, but in contrast to its static counter-part the results only cover a single or few runs. To bridge the gap, and find a promising middle-ground, probabilistic/speculative program analysis frameworks have been proposed. These frameworks can be considered to intersect both by being a static program analysis that uses dynamic information.
We have introduced an data-flow framework based on fuzzy sets that supports such analyses. We solved data-flow problems of use for speculative compilation and showed how our analysis unveils opportunities that previous approaches could not express and reason about. We furthermore showed that our data-flow framework based on fuzzy sets admit mechanisms from fuzzy control theory to enhance the analysis result dynamically allowing for a hybrid analysis framework.

\nocite{*}
\bibliographystyle{eptcs}
\bibliography{qapl}

\medskip
\setcounter{section}{-1}
\section*{Appendix A: Omitted proofs}
\begin{proof}[of Theorem \ref{theoremLipFunc}] \textit{Let $x,y,C,w_i \in [0,1]$, for some $i \in \mathbb{N}$, $f_i(\vec{x}): [0,1]_q^n \mapsto [0,1]_q$ be $1$-Lipschitz and $g_i(\vec{x}): [0,1]_q^n \mapsto [0,1]_q$ be $K_i$-Lipschitz.}
\begin{enumerate}
\item \textit{Functions $x+y$, $x-y$, $xy$, $\min(x,y)$ and $abs(x)$ are $1$-Lipschitz. Constants are $0$-Lipschitz}
Let $b \in [x, x+h]$ for some $0 \leq x \leq x+h \leq 1$:
{ \smaller
\begin{enumerate}
 \item \label{caseAbs} $g(x) = abs\left(x\right)$:
  \begin{align*}
   \left|g(x+h)-g(x)\right| &= \left|abs\left(x+h\right) - abs\left(x\right)\right| && \text{By definition} \\
 	&= \left|x+h - x\right| && x,h \in [0,1] \\
 	&\leq 1|h|
  \end{align*}
 \item \label{caseAdd} $g(x,y) = x + y$:
  \begin{align*}
   \left|g(x+h_1, y+h_2)-g(x, y)\right| &= \left|((x+h_1)+(y+h_2)) - (x+y)\right| && \text{By definition} \\
  	&= |h_1 + h_2| && \\
	&\leq |h_1| + |h_2| && \text{Triangle inequality} \\
    &= 1|h| && \text{Distributivity}
  \end{align*}
 \item \label{caseSub} $g(x,y) = x - y$:
  \begin{align*}
   \left|g(x+h_1, y+h_2)-g(x, y)\right| &= \left|((x+h_1)-(y+h_2)) - (x-y)\right| && \text{By definition} \\
  	&= |h_1 + (-1)h_2| && \\
	&\leq |h_1| + |-1||h_2| && \text{Triangle inequality} \\
    &= 1|h| && \text{Distributivity}
  \end{align*}
 \item \label{caseMul} $g(x,y) = xy$:
  \begin{align*}
   \left|g(x+h_1, y+h_2)-g(x, y)\right| &= \left|g(x+h_1, y')-g(x, y'-h_2)\right| && \text{Substitution }y' = y+h_2 \\
    &= \left|(x+h_1)y' - x(y'-h_2)\right| && \text{By definition} \\
  	&= |h_1y' + xh_2| \\
  	&\leq |h_1 + h_2| && 0 \leq x,y' \leq 1 \\ 
	&\leq |h_1| + |h_2| && \text{Triangle inequality} \\
    &= 1|h| && \text{Distributivity}
  \end{align*}
 \item \label{caseMin} $g(x,y) = \min(x, y)$:
  \begin{align*}
   \left|g(x+h_1, y+h_2)-g(x, y)\right| &=	\left|\min(x+h_1, y+h_2)-\min(x, y)\right| && \text{By definition} \\
 	&= \begin{aligned}
 	      &\left|\left(\frac{x+h_1+y+h_2}{2} - \frac{|x+h_1-y-h_2|}{2}\right)\right. - \\
 	      &\left.\left(\frac{x+y}{2} - \frac{|x-y|}{2}\right)\right|
 	   \end{aligned} && min(x,y) = \frac{x+y}{2} - \frac{|x-y|}{2} \\
 	&\leq \begin{aligned}
 	        &\left|\frac{h_1+h_2}{2} - \frac{|x|+|-1||y|+|h_1-h_2|}{2} +\right. \\
 	        &\left.\frac{|x|+|-1||y|}{2} \right|
 	       \end{aligned}\\
 	&= \left|\frac{h_1+h_2}{2} - \frac{|h_1-h_2|}{2}\right| \\
 	&= |\min(h_1, h_2)| \\
 	&\leq |h_1 + h_2| \\
	&= 1|h| && \text{Triangle inequality}
  \end{align*}
 \item \label{caseConstant} $g(x,y) = C$
 	\begin{align*}
 		|g(x+h) - g(x)| = |C - C| = 0 \leq 0|h|
 	\end{align*}
\end{enumerate}
}

\item \textit{If $\sum_{i = 0}^{N-1} w_i = 1$ then $\sum_{i = 0}^{N-1} w_if_i(\vec{x})$ is $1$-Lipschitz}\label{caseAvg}

\begin{align*}
	\left|g(\vec{x}+\vec{h})- g(\vec{x})\right| &= \left|\sum_{i = 0}^{N-1} w_if(\vec{x+h}) - \sum_{i = 0}^{N-1} w_if(\vec{x})\right| && \text{By definition} \\
	&= \left|\sum_{i = 0}^{N-1} w_i\left(f(\vec{x+h})-f(\vec{x})\right)\right| && \text{Associativitiy and commutativity} \\
	&\leq \left(\sum_{i = 0}^{N-1} w_iK_i\right)|h| && 
		\text{Triangle inequality, distributivity and } w_i \geq 0 \\
	&= 1|h| && K_i = 1\text{ and } \sum_i w_i = 1
\end{align*}

\item \textit{The composition $g_a \circ g_b$ is $K_aK_b$-Lipschitz}\label{caseComp}
	$g(\vec{x}) = f_a(\vec{x}) \circ f_a(\vec{x})$:
  \begin{align*}
   \left|g(x+h)-g(x)\right| &= \left|f_a(f_b(\vec{x+h})) - f_a(f_b(\vec{x}))\right| && \text{By definition} \\
 	&\leq K_a\left|f_b(\vec{x+h}) - f_b(\vec{x})\right| && \text{Definition \ref{defLipFunc}} \\
 	&\leq K_aK_b|h| && \text{Definition \ref{defLipFunc}}
  \end{align*}
 
\item \textit{Formulas defined in a Frank family Fuzzy logic are $1$-Lipschitz.}
This follows from structural induction on the height of parse tree of the predicate $P(x)$. By De Morgan's laws it is enough to show the induction step for $\vee$ and $\lnot$. 
\begin{itemize}
 \item \textbf{Base case}:
  \begin{itemize}
   \item $v$: $g(x) = x$: $|g(x+h) - g(x)| = |x+h-x| \leq 1|h|$.  
   \item $\top$ or $\bot$: $g(x) = C$: $1$-Lipschitz by \ref{caseComp}.  
  \end{itemize}
 \item \textbf{Induction step}:
  \begin{itemize}
   \item $\lnot \phi \equiv 1-\phi$: $1$-Lipschitz from the base case for constants ($\top$ and $\bot$) and cases \ref{caseSub} and \ref{caseComp} and assumption that $\phi$ is $1$-Lipschitz.
   \item $\phi_1 \vee \phi_2$:
     $\left\{\begin{array}{ll}
		\min(x,y) & 1\text{-Lipschitz from Theorem \ref{theoremLipFunc} case \ref{caseMin}}\\
		xy & 1\text{-Lipschitz from Theorem \ref{theoremLipFunc} case \ref{caseMul}} \\
		\max(x+y-1,0) & \text{Equal to }1-\min(2-x-y,1)\text{ using De Morgans law.} \\
		& \text{This expression is }1\text{-Lipschitz from Theorem \ref{theoremLipFunc}} \\
                & \text{ case \ref{caseSub}, \ref{caseMin} and \ref{caseComp}}
   	 \end{array}\right.$
  \end{itemize}
\end{itemize}

\item \textit{If $F: \mathbb{I}_q^n \to \mathbb{I}_q$ satisfy $\forall x \in \mathbb{I}_q^n: y \in x \Rightarrow f(y) \in F(x)$ then F is $1$-Lipschitz}

$F: \mathbb{I}_q^n \to \mathbb{I}_q$ can be decomposed into two functions $F_l: \mathbb{I}_q^n \to [0,1]_q$ and $F_u: \mathbb{I}_q^n \to [0,1]_q$ such that $\forall i: F(i) = [F_l(i), F_u(i)]$, i.e., $F_l(i)$ gives the infimum of $f(i)$ and $F_u(i)$ gives the supremum of $i$. We show that both $F_l$ and $F_u$ are $1$-Lipschitz continuous:
\begin{itemize}
 \item $F_l(I) = \inf\limits_{i \in I} f(i)$: Assume $I = [l, u]$, since $I$ is finite we can rewrite the operation as pair-wise applications of $\min$, i.e., $min(f(l), min(f(l+\frac{1}{2^{q}}), min(f(l+\frac{2}{2^{2}}), ...)))$. As per above case \ref{caseMin} $\min$ is $1$-Lipschitz. Similarly the composition of two $1$-Lipschitz functions is also $1$-Lipschitz, or in extension, a finite number of compositions.
 \item $F_u(I) = \sup\limits_{i \in I} f(i)$: $\max(x,y)$ is equivalent to $1-\min(1-x, 1-y)$ which is $1$-Lipschitz by above case \ref{caseSub} and \ref{caseMin} so proof follows in the same way as the $F_l(I)$ case.
\end{itemize}

\end{enumerate}
\end{proof}

\end{document}